\documentclass[aps,prb,twocolumn,superscriptaddress,showpacs,amsmath,amssymb]{revtex4} 
\usepackage{graphicx}
\usepackage{appendix}
\usepackage{epsfig} 
\usepackage[T1]{fontenc}
\usepackage[english]{babel}

\begin{document}

\newcommand{\io}{i\omega_n}
\newcommand{\cdag}{c^{\dagger}}
\newcommand{\fdag}{f^{\dagger}}
\newcommand{\dagga}{{\phantom{\dagger}}}
\newcommand{\up}{\uparrow}
\newcommand{\dw}{\downarrow}
\newcommand{\m}[1]{\mathcal{#1}}
\newcommand{\mb}[1]{\mathbf{#1}}
\newcommand{\ta}[2]{\tau^{#1}_{#2}}
\newcommand{\tb}[2]{\tau^{'#1}_{#2}}
\newcommand{\al}[2]{\alpha^{#1}_{#2}}
\newcommand{\ab}[2]{\alpha^{'#1}_{#2}}
\newcommand{\x}[1]{x_{#1}^{a}}
\newcommand{\y}[1]{x_{#1}^{'a}}
\newcommand{\cc}[1]{c_{a\ab{a}{#1}}(\tb{a}{#1})}
\newcommand{\cd}[1]{c^{\dag}_{a\al{a}{#1}}(\ta{a}{#1})}
\newcommand{\DD}[1]{\Delta^a_{\al{a}{#1}\ab{a}{#1}}(\ta{a}{#1},\tb{a}{#1})}
\newcommand{\DDD}[2]{\Delta^a_{\al{a}{#1}\ab{a}{#2}}(\ta{a}{#1},\tb{a}{#2})}
\newcommand{\f}[1]{\m{#1}\left(\m{C}\right)}
\newcommand{\eps}{\varepsilon}
\newcommand{\eqn}[1]{(\ref{#1})}
\newcommand{\be}{\begin{equation}}
\newcommand{\ee}{\end{equation}}
\newcommand{\bea}{\begin{eqnarray}}
\newcommand{\eea}{\end{eqnarray}}
\newcommand{\ba}{\begin{eqnarray*}}
\newcommand{\ea}{\end{eqnarray*}}

\title{Real-Time Dynamics in Quantum Impurity Models with Diagrammatic Monte Carlo}
\author{Marco Schir\'o}
\affiliation{International School for Advanced Studies (SISSA), and CRS Democritos, CNR-INFM,
Via Beirut 2-4, I-34014 Trieste, Italy} 
\date{\today} 

\begin{abstract}

We extend the recently developed real-time Diagrammatic Monte Carlo method, in its hybridization expansion formulation, to the full Kadanoff-Baym-Keldysh contour. This allows us to study real-time dynamics in correlated impurity models starting from an arbitrary, even interacting, initial density matrix.
As a proof of concept we apply the algorithm to study the non equilibrium dynamics after a local quantum quench in the Anderson Impurity Model. Being a completely general approach to real-time dynamics in quantum impurity models it can be used as a solver for Non Equilibrium Dynamical Mean Field Theory. 

\end{abstract}
\maketitle

\section{Introduction}

The understanding of real-time dynamics in strongly correlated quantum systems represents a major challenge in modern condensed matter physics, 
due to the rapid experimental advances in probing physical responses directly in the time domain\cite{Cavalieri_nature06}. 
Time resolved pump-probe experiments with femtosecond resolution, for example, have been recently moving from the realm of atoms 
and molecules\cite{Zewail_jchem_00} to that of strongly correlated bulk materials, such as Mott 
Insulators\cite{Perfetti_Georges_PRL06} and High-Temperature Superconductors\cite{Giannetti_PRB09_BISCCO}, 
thus opening the way toward a full characterization of their non equilibrium properties. Similarly, the emerging field of 
cold atomic quantum gases represents a natural laboratory where the dynamics of almost isolated quantum systems can be probed in 
real-time. Several experiments have been performed in this direction to measure, for instance, the dynamics after a 
quantum quench in bosonic gases\cite{Greiner_nature_02}, the lifetime of doublons in strongly interacting Fermi 
systems\cite{Doublon_decay} or the onset of superexchange interactions between ultra cold atoms in optical 
lattices\cite{Superexchange_science}. Finally, we note that even at the nanoscale level the experimental frontiers are 
rapidly moving in the direction of time-resolved techiniques to detect charge transport by counting individual electrons 
while tunneling across correlated nanostructures such as semiconducting quantum 
dots\cite{Elzerman_APL_04,Ensslin_APL_04,Gustavsson_2009}. 

From the theoretical point of view, the field of real-time dynamics in strongly correlated systems is still at its infancy. 
In this perspective Quantum Impurity (QI) models
represent the ideal playground to test and develop new methods. These models consist of a small quantum system 
with few interacting degrees of freedom, the impurity,  coupled via hybridization to a reservoir of fermionic excitations. 
QI models therefore represent, by construction, the natural framework to study quantum transport through nanocontacts.
While the equilibrium physics of these nutshell strongly correlated systems can be studied with a wide range of powerful numerical 
and analitical tools, their non equilibrium real-time dynamics is still challenging. The reason for this gap is mainly due 
to the fact that most of the theoretical tools which has been developed in the last thirty years to solve quantum impurity 
models in equilibrium, most notably Numerical Renormalization Group\cite{Wilson_RMP,Pruschke_RMP} (NRG), can not be directly 
applied to the out of equilibrium case. 
This has triggered a large amount of theoretical works among which we mention the time dependent 
extensions of NRG\cite{tnrg_Anders} and DMRG\cite{DMRG_Dagotto_09,Weichselbaum_PRB_09}, the ISPI method\cite{iter_Egger} 
and the Flow equation approach\cite{Kehrein_PRB_08}.\\
Beside their relevance for nanoscience, QI models have also been emerging in the last two decades as the paradigm to 
understand strong correlation phenomena in bulk lattice models within the so called Dynamical Mean Field 
Theory\cite{Review_DMFT_96}. The 
extension of this powerful non perturbative technique to the non equilibrium 
domain\cite{Non_Eq_DMFT,Werner_DMFT} makes the development of a generic numerically exact approach to real-time dynamics in 
quantum impurities an even more urgent issue.

Recently, a new generation of Diagrammatic Monte Carlo (diagMC) algorithms have been developed to solve 
equilibrium quantum impurity models. These are based on a stochastic sampling of the partition function written 
as an imaginary-time diagrammatic expansion around weak or strong coupling values of the interaction acting on the 
impurity\cite{ctqmc_Rubtsov,ctqmc_Werner}. 
An extension to the out of equilibrium case, namely to the real-time domain, has been proposed only very recently. 
The general idea of Real-Time diagMC methods is to start from a given initial density matrix, describing the fermionic 
reservoir and the impurity, and to compute the dynamics of any observable of the system by sampling the real-time evolution 
operator written as a diagrammatic expansion along the Keldysh contour. Both the weak and the strong coupling expansion 
have been proposed and applied to the local polaron problem\cite{Keldysh_short,Rabani} and to the 
Anderson impurity model\cite{Werner_Keldysh_09}. 
By construction, these approaches rely on a rather specific assumption on the initial density matrix, which has to describe 
either a non-interacting impurity model or an impurity decoupled from the reservoir. 
Going beyond this assumption, which is technical rather than fundamental, is the main motivation for this work. 
To this extent we merge together the imaginary time and the real-time methods thus developing a completely general Diagrammatic 
Monte Carlo algorithm on the full three-branches Kadanoff-Baym-Keldysh contour\cite{Kadanoff_Baym,Rammer_book_noneq}. 
This allows us to deal with the most generic non equilibrium set-up, namely an \emph{interacting} quantum impurity model described 
at time $t=0$ by some thermal density matrix, driven out of equilibrium for time $t>0$ by a generic, possibly time dependent, perturbation.
As a by product, we obtain a general real-time solver for quantum impurity models which can be therefore used to solve 
Non Equilibrium Dynamical Mean Field Theory. Even in this context, the possibility of studying the non equilibrium real-time dynamics starting from interacting initial states look particularly intriguing in the light of first DMFT results on quantum quenches in the Hubbard model\cite{Werner_DMFT} which focus so far only on quenches starting from a non interacting initial state.

As a first application of this algorithm we study the non equilibrium real-time dynamics in the Anderson Impurity model after a local quantum quench. 

The paper is organized as follows. In section II we introduce the formalism. 
In section III we derive the hybridization expansion on the Kadanoff-Baym-Keldysh contour while section IV is devoted to the description of the diagMC algorithm. Section V describes the application to the Anderson Impurity Model while section VI is for conclusions and perspectives.

\section{Non Equilibrium Dynamics in Quantum Impurity Models}

The aim of this section is to set-up the proper framework to study non equilibrium real-time dynamics in quantum impurity models. 
To this purpose, we consider a set of discrete electronic levels, the ``impurity'', with creation operator $\cdag_{a}$ 
labeled by an integer $a=1,\dots\m{N}$ which may include both spin and orbital degrees of freedom. These levels are coupled to 
one or more baths of free fermions with momentum $\mathbf{k}$ and creation operator $\fdag_{\mathbf{k}\,a}$. 
The generic hamiltonian of a QI model reads
\bea\label{eqn:H_-}
\m{H}_- & = &\sum_{\mathbf{k}\,a}\,\varepsilon^-_{\mathbf{k}}\,
\fdag_{\mathbf{k}\,a}\,f_{\mathbf{k}\,a}+
\m{H}^-_{loc}\left[\cdag_{a},c_{a}\right]\nonumber\\
&&+\sum_{\mathbf{k}\,a}\left(V^-_{\mathbf{k}\,a}\,\fdag_{\mathbf{k}\,a}\,c_{a} + h.c.\right)\,,
\eea
where the first term describes the continuum of fermionic excitations, the local hamiltonian 
$\m{H}^{-}_{loc}\left[\cdag_{a},c_{a}\right]$ generally contains many-body interactions 
for electrons on the impurity, while the last term is the hybridization which couples the impurity and the bath and it is 
assumed here, for the sake of simplicity, to be diagonal in the $a$ index.

Since we are interested in studying non equilibrium dynamics of model ~(\ref{eqn:H_-}), we have to specify an initial condition 
as well as a protocol to drive this system out of equilibrium. Following general ideas of non equilibrium many body 
theory\cite{Rammer_book_noneq,Kadanoff_Baym}, we imagine to prepare the system at $t=0$ in a thermal state of $\m{H}_-$, 
namely we choose the Boltzmann distribution as initial density matrix 
\be\label{eqn:rho_therm}
\rho(t=0)=\rho_{eq}\equiv \frac{e^{-\beta \m{H}_-}}{Z}\,,\qquad Z = \mbox{Tr}\,e^{-\beta \m{H}_-}\,,
\ee
and then, for $t>0$, let the system evolve under the action of a new hamiltonian
\be\label{eqn:H_evolution_general}
\m{H}\left(t\right) = \m{H}_{-} + \m{V}\left(t\right)\,,\qquad t>0
\ee
Choosing the initial density matrix as the thermal one gives access to the response of 
a \emph{correlated} quantum impurity model to external fields. As we shall discuss in what follows, 
Eq.~(\ref{eqn:rho_therm}) represents the main point where our approach differs from previous implementations 
of the real-time diagrammatic Monte Carlo method\cite{Keldysh_short,Rabani,Werner_Keldysh_09}. For what concerns the driving 
protocol, namely the nature of the external perturbation, there are actually several ways to push a quantum impurity model 
out of equilibrium. In this work we shall focus on the simplest one, namely a quantum quench experiment, but the method 
allows to address even more general \emph{time dependent} out of equilibrium problems.

In a quantum quench, one imagines to prepare the system, at $t=0$, in a given state of some initial hamiltonian 
($\m{H}_-$ in the case of our interest) and then, for $t>0$, to suddenly change some of its parameters letting evolve the system 
under the unitary action of a new Hamiltonian $\m{H}_+$. Such a protocol therefore represents the simplest example of 
time-dependent problem where the variation in time is step-like
\be\label{eqn:H_t}
\m{H}\left(t\right) = \m{H}_{-} + \theta\left(t\right)\delta\,\m{H}\,,\qquad
\delta\m{H} = \m{H}_+ - \m{H}_-\,.
\ee
The sudden quench injects energy into the system and leads to a relaxation dynamics towards a new steady state, 
provided the perturbation $\delta\m{H}$ is not a conserved 
quantity of $\m{H}_-$. The main task is therefore to compute quantum averages with the full density matrix 
$\rho(t)$ evolved in real-time
\begin{equation}\label{eqn:O_aver}
 \langle O(t)\rangle = Tr\,\left[\rho\left(t\right)\,\m{O}\,\right]\,=
Tr\,\left[\rho_{eq}\,U^{\dagger}\left(t\right)\,\m{O}\,U\left(t\right)\right]\,.
\end{equation}
where the trace has to be taken over the bath and the impurity degrees of freedom, while $U\left(t\right)$ 
and $U^{\dagger}\left(t\right)$ are, respectively, the unitary operator generating the dynamics and its hermitian conjugate. 
In the specific case of a time independent hamiltonian, as we have for $t>0$ see Eq.~(\ref{eqn:H_t}), these operators read
\be
U\left(t\right) = e^{-i\,\m{H}_{+}\,t}\,\qquad
U^{\dagger}\left(t\right) = e^{i\,\m{H}_{+}\,t}\,.
\ee
To proceed further, it is convenient to specify the nature of the perturbation $\delta\,\m{H}$ induced by the quantum quench. 
To keep the discussion as general as possible we write the hamiltonian of the system for $t>0$ as
\bea\label{eqn:H_+}
\m{H}_{+} & = &\sum_{\mathbf{k}\,a}\,\varepsilon^{+}_{\mathbf{k}}\,\fdag_{\mathbf{k}\,a}\,f_{\mathbf{k}\,a}+
\m{H}^{+}_{loc}\left[\cdag_a,c_a\right]\nonumber\\
&&+\sum_{\mathbf{k}\,a}\left(V^+_{\mathbf{k}\,a}\,\fdag_{\mathbf{k}\,a}\,c_a + h.c.\right)\,,
\eea
namely we allow for an abrupt change of all the parameters entering in the hamiltonian (\ref{eqn:H_-}), in such a way that 
different kind of non equilibrium phenomena can be treated within the present approach. One can, for example, study the dynamics 
after a local quantum quench acting only on the impurity degrees of freedom, or considering a global change in the hamiltonian, 
acting on the hybridization term or even on the conduction band. These global quantum quenches are particularly relevant for 
studying dc transport in quantum dots and will be the subject of a forthcoming publication. 

Once we have specified the structure of the hamiltonian after the quench, we can perfom the hybridization expansion in complete 
analogy to what has been done previously in the case of a pure Keldysh real-time 
algorithm\cite{Keldysh_short,Rabani,Werner_Keldysh_09} with, nonetheless, an important difference. 
Indeed, as it appears clearly from the above formulation (see Eq.(\ref{eqn:O_aver}), not only the real-time evolution but also the 
``thermal'' one is governed by the full hamiltonian $\m{H}_{\pm}$, involving both the impurity and the bath degrees of freedom 
coupled one to each other. This allows us to overcome the limitation of the pure Keldysh approach which relies on the assumption of 
an initially decoupled density matrix and gives us the possibility to treat arbitrary initial conditions. In the next sections 
we will describe the Diagrammatic Monte Carlo algorithm used to compute the real-time average in Eq.~(\ref{eqn:O_aver}).

\section{Hybridization Expansion on the Kadanoff-Baym-Keldysh Contour}\label{sect:hybrid_exp}

In order to study the non equilibrium real-time dynamics of quantum impurity models starting from a generic initial density matrix, 
we formulate the diagrammatic monte carlo algorithm (diagMC), in its hybridization expansion version, on the 
Kadanoff-Baym-Keldysh contour made by the usual imaginary time axis and the real-time Keldysh contour. 
As we are going to show, this structure naturally emerges from the definition of real-time quantum averages given in 
equation (\ref{eqn:O_aver}). To proceed further, we introduce a \emph{dynamical} time-dependent partition function for the QI model which is 
defined as
\be\label{eqn:Z_dyn_def}
\m{Z}\left(t,\beta\right) \equiv 
\mbox{Tr}\,\left[e^{-\beta \m{H}_{0}}\,U^{\dagger}\left(t\right)U\left(t\right)\right]\,.
\ee
We note that this quantity does not actually depend on time $t$ since, by construction, the evolution is unitary, nevertheless 
it represents the natural quantity to derive the hybridization expansion. As it will appear more clearly later on, 
$\m{Z}\left(t,\beta\right)$ can be seen a dynamical generating functional of the Monte Carlo weights needed to compute any 
quantum average in real-time.
The basis of any continuous-time diagMC algorithm is the expression of evolution operators as time-ordered exponentials. 
For the real-time operator and its hermitian conjugate we get
\bea\label{eqn:time_ord_real}
U\left(t\right) & = & \mbox{T}\,\exp\left(-i\int_{0}^{t}\,dt\,\m{H}_+\left(t\right)\right)\\
U^{\dagger}\left(t\right) & = & \bar{\mbox{T}}\,\exp\left(i\int_{0}^{t}\,dt\,\m{H}_+\left(t\right)\right)
\eea
where $\mbox{T}$ ($\bar{\mbox{T}}$) is the time ordering (anti-time ordering) operator, whose action is order a 
string of real-time fermionic operators according to their time arguments, placing to the left the operators  
at later (earlier) times, with an overall plus or minus sign according, respectively, to the parity of the number of fermionic 
exchanges needed to move the string from the original to the final ordering.
Using the well known properties of the equilibrium density matrix (\ref{eqn:rho_therm}) we can write also the Boltzmann weight 
as an imaginary time evolution along the path $\left[-i\beta,0\right]$
\bea\label{eqn:time_ord_imag}
e^{-\beta\m{H}_-} &= &\mbox{T}\,\exp\left(-\int_{0}^{\beta}\,d\tau\,\m{H}_-\left(-i\tau\right)\right)\\
&=& \mbox{T}_{\rceil}\,\exp\left(-i\int_0^{-i\beta}\,dt\,\m{H}_-\left(t\right)\right) = U\left(-i\beta\right)\nonumber\,,
\eea
where $\mbox{T}_{\rceil}$ is an imaginary-time-ordering operator defined in complete analogy with $\mbox{T}$.

\begin{figure}[ht]
\begin{center}
\psfig{figure=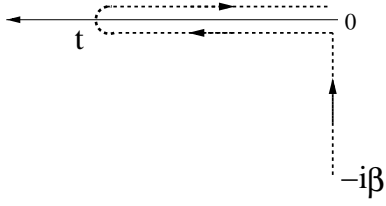,scale=0.35}
\caption{Kadanoff-Baym-Keldysh contour which starts at time $0$ on the first branch, runs up to time $t=t_{\ast}$ then back from $t_{\ast}$ to $t=0$ and finally along the imaginary axis from $t=0$ to $t=-i\beta$.}\label{fig:KBK} 
\end{center}
\end{figure}

Inserting these expressions in the dynamical partition function previously introduced, we get
\bea\label{eqn:Z_dyn_contour}
\m{Z}\left(t\right) & = & \mbox{Tr}\,\left[
\mbox{T}_{\rceil}\,e^{i\int_{-i\beta}^0\,dt\,\m{H}_-\left(t\right)}\;
\bar{\mbox{T}}\,e^{i\int_{0}^t\,dt\,\m{H}_+\left(t\right)}\;
\mbox{T}\,e^{i\int_{t}^0\,dt\,\m{H}_+\left(t\right)}\right]\nonumber\\
&=& \mbox{Tr}\,\left[\mbox{T}_{C}\,e^{i\int_{C}\,dt\,\m{H}\left(t\right)}\right]
\eea
where, in the second line, $C$ is the Kadanoff-Baym-Keldysh contour plotted in figure \ref{fig:KBK}, which is made of 
three branches $\m{B}_i$, $i=1,2,3$, being respectively the upper real-time branch, the lower one and the imaginary-time branch. 
Hereafter the time argument $t$ in Eq.~(\ref{eqn:Z_dyn_contour}) is assumed to live on such a contour, unless differently specified.  
Time ordering along $C$, enforced by operator $\mbox{T}_C$, acts similarly to the standard Keldysh time-ordering, 
placing operators with later time on the left, namely 
\be
T_C\left(A\left(t_1\right)A\left(t_2\right)\right)=
\left\{
\begin{array}{cc}
A\left(t_1\right)A\left(t_2\right) &  t_1\stackrel{C}{>} t_2 \\
&\\
-A\left(t_2\right)A\left(t_1\right) & t_1\stackrel{C}{<} t_2
\end{array}
\right.\,,
\ee
where $\stackrel{C}{>}$ ($\stackrel{C}{<}$) means greater (lesser) on the contour $C$.
By the definition of the partition function $\m{Z}\left(t\right)$, it follows that 
\be
\mbox{T}_C = \mbox{T}_{\rceil}\otimes\bar{\mbox{T}}\otimes\mbox{T}\,.
\ee
The integral along the contour is defined as
\be\label{eqn:contour_int}
\int_C dt = \int_{-i\beta}^0 dt + \int_0^{t}dt+ \int_t^{0}dt\,,
\ee
while, along the contour, the Hamiltonian entering in Eq.~(\ref{eqn:Z_dyn_contour}) is 
\be\label{eqn:contour_H}
\m{H}\left(t\right)=
\left\{
\begin{array}{cc}
t\in\m{B}_1,\m{B}_2 &  \qquad\m{H}_+\\
t\in\m{B}_3 &  \qquad\m{H}_-
\end{array}
\right.
\ee
Once we have written the partition function $\m{Z}\left(t,\beta\right)$ as a time ordered exponential, 
we can treat different terms in the contour hamiltonian as c-numbers. Therefore we can write Eq.(\ref{eqn:Z_dyn_contour}) as
\be\label{eqn:Z_dyn_hyb_exp_0}
\m{Z}\left(t,\beta\right)=
\mbox{Tr}\,\left[\mbox{T}_{C}\,e^{i\int_{C}\,dt\,\m{H}_0\left(t\right)+\m{H}_{hyb}\left(t\right)}\right]\,
\ee
where we have explicitly indicated the hybridization Hamiltonian $\m{H}_{hyb}\left(t\right)$, defined as
\be\label{eqn:H_hyb}
\m{H}_{hyb}\left(t\right)=
\sum_{\mathbf{k}\,a}\left(V_{\mathbf{k}\,a}\left(t\right)\,\fdag_{\mathbf{k}\,a}\left(t\right)
\,c_a\left(t\right) + h.c.\right)\,
\ee
with a time dependent hybridization $V_{\mathbf{k}\,a}\left(t\right)= V^{\pm}_{\mathbf{k}\,a}$ depending on the position of 
time $t$ along the contour $C$. To proceed further, it is convenient to introduce the bath operators at the impurity site, 
defined as
\be\label{eqn:bath_ope_loc}
\fdag_a\left(t\right) = \sum_{\mathbf{k}}\,V_{\mathbf{k}\,a}\left(t\right)\,\fdag_{\mathbf{k}\,a}\left(t\right)\,
\ee
which enter the hybridization Hamiltonian Eq.(\ref{eqn:H_hyb}). Then we formally expand $\m{Z}\left(t,\beta\right)$ in power 
of the hybridization Hamiltonian (\ref{eqn:H_hyb}) and trace out, at any order in the expansion, separately the bath and the impurity 
degrees of freedom which are completely decoupled in the absence of $\m{H}_{hyb}$. Let us define $n_{ab}$  ($\tilde{n}_{ab}$)
as the number of creation(annihilation) impurity operators, in the following also called kinks, with flavour $a$ and in branch $b$. 
These integers run in general between zero and infinity, the only constraint is that the total number of creation and annihilation 
kinks for each flavour $a$, say $k_a$ and $\tilde{k}_a$, have to be equal due to total particles conservation. This introduces 
the following constraint
$$
\tilde{k}_a\equiv\sum_b\,\tilde{n}_{ab}\equiv k_a \equiv \sum_b\,n_{ab}\,\qquad a=1,\dots,\m{N}
$$
The resulting expansion for the partition function reads
\begin{widetext}
\begin{eqnarray}\label{eqn:Z_dyn_hyb_exp_1}
\m{Z}\left(t,\beta\right) &=&
\prod_{a=1}^{\m{N}}\,
\prod_{b=1}^{3}\,
\sum_{\tilde{n}_{ab},n_{ab}}\,
\mbox{s}(\tilde{n}_{ab},n_{ab})\;
\prod_i^{k_a}\,
\int dt^a_i\,
\int d\tilde{t}^a_i\;
\prod_a\,\langle\,T_{C}\,
f\left(t^a_1\right)f^{\dagger}\left(\tilde{t}^a_1\right)\dots f\left(t^a_{k_a}\right)f^{\dagger}
\left(\tilde{t}^a_{k_a}\right)\rangle_{bath}\;\nonumber\\
& &\qquad\qquad\qquad\qquad\qquad
\times
\langle\,T_{C}\,\prod_a\left(c^{\dagger}\left(t^a_1\right)c\left(\tilde{t}^a_1\right)\dots c^{\dagger}
\left(t^a_{k_a}\right)c\left(\tilde{t}^a_{k_a}\right)\right)\,
\rangle_{local}\,,
\end{eqnarray}
\end{widetext}
where the contour integrals must be constrained to the time ordered regions
\be\label{eqn:constraint}
t_1^a \stackrel{C}{>}t_2^a\stackrel{C}{>}\ldots\stackrel{C}{>}t^a_{k_a}\qquad,\qquad
\tilde{t}_1^a\stackrel{C}{>}\tilde{t}_2^a\stackrel{C}{>}\ldots\stackrel{C}{>}\tilde{t}^a_{k_a}\,,
\ee
while $\mbox{s}(\tilde{n}_{ab},n_{ab})$ includes all the signs(phases) coming from the real-time evolution operators as well as from the integration 
along the imaginary time axis
\be\label{eqn:signs}
\mbox{s}(\tilde{n}_{ab},n_{ab})=
(-i)^{\tilde{n}_{a3}+n_{a3}}\,
i^{\tilde{n}_{a2}+n_{a2}}\,
(-i)^{\tilde{n}_{1a}+n_{1a}}\,.
\ee
By using our definition of the total number of kinks $k$ we can write these factors as
\be\label{eqn:signs_2}
\mbox{s}(\tilde{n}_{ab},n_{ab})=(-1)^{k_a}\,(-1)^{\tilde{n}_{a2}+n_{a2}}\,
(-i)^{\tilde{n}_{a3}+n_{a3}}\,.
\ee
Concerning the trace over the bath degrees of freedom with flavour $a$, this can be written by using Wick's theorem 
as the determinant of a $k_a\times k_a$ matrix $\mathbf{\Delta}^a$
\be\label{eqn:det}
\langle\,T_{C}\,
f\left(t^a_1\right)f^{\dagger}\left(\tilde{t}^a_1\right)\dots f\left(t^a_{k_a}\right)f^{\dagger}
\left(\tilde{t}^a_{k_a}\right)\rangle_{bath}=
\mbox{det}\mathbf{\Delta}^a
\ee
whose entries $\Delta^a_{ij}$ are the contour-ordered hybridization functions defined as
\bea\label{eqn:delta_contour}
\Delta^a_{ij}&\equiv& i\Delta_C\left(t^a_i,\tilde{t}^a_j\right)=\nonumber\\
&=&\sum_{\mathbf{k}}\,V_{\mathbf{k}\,a}\left(t^a_i\right)\,
\langle\,T_C\,f_{\mathbf{k}}\left(t^a_i\right)
f^{\dagger}_{\mathbf{k}}\left(\tilde{t}^a_j\right)\rangle\,
V_{\mathbf{k}\,a}\left(\tilde{t}^a_j\right)\,
\eea
the average being taken over the bath degrees of freedom. 
This function is defined along the Kadanoff-Baym-Keldysh contour and therefore naturally acquires the structure of 
a $3\times3$ matrix in the branch space\cite{Wagner}, as we discuss in the appendix \ref{app:contour_ord_hyb}.

Unlike what happens for the bath, the trace over the local hilbert space cannot be written in terms of single particle 
contractions since in general Wick's theorem does not hold for an interacting impurity. This consideration holds regardless 
of the specific form of the local hamiltonian $\m{H}_{loc}$ and it is also true in equilibrium. It is therefore clear that the evaluation of the local term in Eq.~(\ref{eqn:Z_dyn_hyb_exp_1}) for a given configuration of kinks is the computational bottlneck of the algorithm, expecially in the case where multi-orbital interactions are considered. We note that such a kind of expressions arise also in NCA-kind of approaches to QI models\cite{Keiter_84}. To efficiently evaluate this local term  we follow\cite{Werner_ctqmc_matrix,Haule_ctqmc} and 
take advantage of the reduced hilbert space of the impurity to write the multi-point correlation function of 
Eq.~(\ref{eqn:Z_dyn_hyb_exp_1}) in the basis of the local eigenstates.
The trace then reduces to multiplying matrices sandwiched by local evolution operators. If we define a global time ordering 
along the contour such that
$$
t_1>t_2>\dots>t_{2N}\,,
$$
where $N=\sum_{a}\,k_a=\sum_{a}\,\tilde{k}_a$, then the local trace can be written as
\bea\label{eqn:loc_trace}
&
\langle\,T_{C}\,\prod_a\left(c^{\dagger}\left(t^a_1\right)c\left(\tilde{t}^a_1\right)\dots c^{\dagger}
\left(t^a_{k_a}\right)c\left(\tilde{t}^a_{k_a}\right)\right)\,
\rangle_{local}
=&\nonumber\\
&s_{T_C}\,\mbox{Tr}\left[\rho_{loc}\,X_1\left(t_1\right)\,X_2\left(t_1\right)\,\dots\,
X_N\left(t_N\right)\,\right]&,
\eea
where we have introduced an extra sign $s_{T_C}$ due to time ordering. 
In the previous equation the $X$'s are creation or annihilation operators (depending on the time ordering) evolved 
in time with the local hamiltonian
\be\label{eqn:local_ope}
X_l\left(t_l\right)=\,
e^{i\,\m{H}_{loc}t_l}\,X_l\,
e^{-i\,\m{H}_{loc}t_l}
\qquad l=1,\dots,2N\,,
\ee
Combining all the above results, we write the hybridization expansion for the dynamical partition function as 
\begin{widetext}
\begin{eqnarray}\label{eqn:hybrid_exp}
\m{Z}\left(t,\beta\right)=
\prod_{a=1}^{\m{N}}\,
\prod_{b=1}^{3}\,
\sum_{\tilde{n}_{ab},n_{ab}}
\,\mbox{s}(\tilde{n}_{ab},n_{ab})\;
\prod_i^{k_a}\,
\int^{'}dt^a_i\,
\int^{'}d\tilde{t}^a_i\;
\prod_a\,\mbox{Det}\,\left[\mathbf{\Delta}^a\right]\,s_{T_C}\,
\mbox{Tr}\left[\rho_{loc}\,X_1\left(t_1\right)\,\,\dots\,X_{2N}\left(t_{2N}\right)\right]
\end{eqnarray}
\end{widetext}
It is worth noticing that this expression represents an exact result for the partition function of the original quantum impurity model. 
As we are going to show in the next section, the goal of the Diagrammatic Monte Carlo method is to sum up 
\emph{stochastically} the hybridization expansion using a Metropolis algorithm. 


\subsection{Effective Action Formulation and Non-Equilibrium DMFT}

An important outcome of the previous calculation has been to obtain an exact and closed expression for the dynamical partition 
function $\m{Z}\left(t,\beta\right)$ of the quantum impurity model, written as a \emph{functional} of the contour-ordered hybridization 
$\Delta_C\left(t,t'\right)$. This result holds in general. 
Indeed, right the same expression for $\m{Z}\left(t,\beta\right)$ is recovered within a path integral formulation of the problem. 
Following Kamenev\cite{Kamenev}, we define the dynamical partition function $\m{Z}\left(t\right)$ as a path integral 
over the fermionic coherent states $c\left(t\right),\bar{c}\left(t\right)$ defined along the three branch contour
\be\label{eqn:Z_dyn_path}
\m{Z}\left(t\right)= 
\int\prod_a\mathcal{D}\bar{c}_a\,\mathcal{D}c_a\;
 e^{i\m{S}_{eff}\left[c_a,\bar{c}_a\right]}
\ee
The effective action describing the real-time dynamics of the impurity model can be written generally as
\be\label{eqn:effective_action}
i\,\m{S}_{eff} =  i\,\m{S}_{loc}+
\int_C\,dt\int_C\,dt'\,\sum_a\,c_a\left(t'\right)i\Delta^a_C\left(t,t'\right)\bar{c}_a\left(t\right)
\ee
where $\m{S}_{loc}$ is the local-in-time impurity action while the quadratic part is defined in terms of the contour-ordered 
hybridization function $i\Delta_C\left(t,t'\right)$ which takes into account the coupling to the bath. 
It is worth noticing here that, by construction, time arguments lie along the three-branch contour $\m{C}$, 
while the integrals are defined as in Eq. (\ref{eqn:contour_int}). As a consequence, the contour-ordered hybridization 
function naturally acquires a $3\times3$ matrix structure in the Keldysh-Matsubara space,
\be\label{eqn:delta_contour_3x3}
i\Delta_C\left(t,t'\right)\rightarrow\;i\Delta_{\alpha\beta}\left(t,t'\right)
\,\qquad \alpha,\beta=1,2,3
\ee
where the $2 \times 2$ block with $\alpha,\beta\neq3$ is the Keldysh subspace, the last diagonal element $\alpha=\beta=3$ is the 
Matsubara sector while the remaining off-diagonal terms are mixed hybridization functions describing the effect of the initial 
condition. We want to show now how is possible within this framework to recover the result for hybridization expansion we 
previously derived. The idea is to formally expand the effective action in power of the contour-ordered hybridization $i\Delta_C$ 
and use the definition of path-integral as the contour time-ordered average of field-operators\cite{Kamenev} to get for 
the partition function $\m{Z}\left(t\right)$ the following expansion
\begin{widetext}
\be
\m{Z}\left(t\right)=\prod_a\,\sum_{k_a}\frac{\left(-1\right)^{k_a}}{k_a!}\int_\m{C}dt_1^a\,d\tilde{t}^a_1
\dots\int_C dt^a_{k_a}\,d\tilde{t}_{k_a}
\;i\Delta^a_C(t^a_1,\tilde{t}^a_1)\dots i\Delta^a_C(t^a_{k_a},\tilde{t}^a_{k_a})\;
\langle\,T_{C}\,\prod_a\left(c^{\dagger}\left(t^a_1\right)c\left(\tilde{t}^a_1\right)\dots c^{\dagger}\left(t^a_{k_a}\right)c\left(\tilde{t}^a_{k_a}\right)\right)\,
\ee
\end{widetext}
where the average is taken over the initial local density matrix. To proceed, we symmetrize the integrand with respect to 
the $k_a!$ permutations of creation times $\left\{t^a_1,t^a_2,\dots,t^a_{k_a}\right\}$ resulting into an extra $1/k_a!$ 
and a determinantal combination of the $k_a$ hybridization functions, where the correct signs to build the determinant are 
provided by the contour time ordered local trace. The $k_a$-th order term in the expansion therefore reads
\bea
&\prod_a\,\frac{\left(-1\right)^{k_a}}{\left(k_a!\right)^2}\int_\m{C}dt_1^a\,d\tilde{t}^a_1
\dots\int_C dt^a_{k_a}\,d\tilde{t}_{k_a}\,
\mbox{Det}\,\left[\mathbf{\Delta}^a\right]\times&\nonumber\\
&\times\langle\,T_{C}\,\prod_a\left(c^{\dagger}\left(t^a_1\right)c\left(\tilde{t}^a_1\right)\dots c^{\dagger}\left(t^a_{k_a}\right)c\left(\tilde{t}^a_{k_a}\right)\right)\, .
&
\eea
Now the integrand is fully symmetric under permutations and we can take advantage of this fact to reduce the size of the 
integration domain, namely we choose among the $k_a!$ possible contour time-orderings for the creation and annihilation times 
with flavour $a$ the following ones
\be
t_1^a \stackrel{C}{>}t_2^a\stackrel{C}{>}\ldots\stackrel{C}{>}t^a_{k_a}\qquad,\qquad
\tilde{t}_1^a\stackrel{C}{>}\tilde{t}_2^a\stackrel{C}{>}\ldots\stackrel{C}{>}\tilde{t}^a_{k_a}\,.
\ee
As anticipated, the final result we get for the dynamical partion function $\m{Z}\left(t\right)$ coincides with the one quoted in Eq. (\ref{eqn:hybrid_exp}). 

These considerations are relevant for studying quantum quenches and real-time dynamics of correlated lattice models within  
Dynamical Mean Field Theory (DMFT). In this case, as one can show explicitly, the full non equilibrium many body problem is mapped, 
in the limit of infinite lattice coordination, onto a quantum impurity model coupled to a \emph{non-equilibrium} bath and 
subject to a self-consistency condition. The dynamical partion function of this effective non equilibrium problem acquires exactly 
the same form as in Eq. (\ref{eqn:Z_dyn_path}), with an unknown contour ordered hybridization function $i\Delta_C\left(t,t'\right)$, 
which generally lacks time traslational invariance. This fact makes uneffective most of the conventional impurity solvers used in 
equilibrium DMFT, which rely on a time independent hamiltonian formulation of the effective problem, thus suggesting diagrammatic 
Monte Carlo method as a natural candidate to solve Non-Equilibrium Dynamical Mean Field equations.

\section{Diagrammatic Monte Carlo }\label{sect:diagMC}
 
Diagrammatic Monte Carlo is a numerical algorithm for sampling infinite series of multiple integrals, 
such as those arising in any perturbative expansion\cite{Prokofev_Svistunov_diagMC_98}. 
As it is well known in many body theory, these expansions often admit a diagrammatic representation\cite{ADG}, even 
in out-of-equilibrium situations. 
This is true also for the hybridization expansion of section~\ref{sect:hybrid_exp}, as we are now going to show,  
which is obtained by a proper extension of the graphical representation  
introduced by Werner and coworkers~\cite{ctqmc_Werner,Werner_ctqmc_matrix} in the context of the imaginary-time 
diagMC algorithm. 

\begin{figure}[ht]
\begin{center}
\psfig{figure=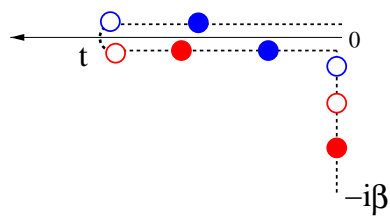,scale=0.35}
\caption{An example of configuration $\m{C}$ sampled by diagMC algorithm.}
\label{fig:config_example} 
\end{center}
\end{figure}

As it can be immediately read out from Eq.~(\ref{eqn:hybrid_exp}), the dynamical partition function can be written as a 
weighted sum over configurations $\m{C}$ made by diagrams on the Kadanoff-Baym-Keldysh contour
\be\label{eqn:Z_dyn_diagMC}
\m{Z}\left(t,\beta\right)=\sum_{\m{C}}\,W\left(\m{C}\right)\,.
\ee
A given configuration contains, for each channel $a=1,\dots N$, a total of $2k_a$ vertices occurring at 
times $\{t^a_i\,,\tilde{t}^a_i\}$ on the contour, with $i=1,\dots k_a$. Half of these vertices represent an impurity creation operator 
$c^{\dagger}_a\left(t^a_i\right)$ while the other half stems for an impurity annihilation operator 
$c_a\left(t^a_i\right)$, both of them being evolved in time with the local hamiltonian $\m{H}_{loc}$.
All together we have $2\sum_a\,k_a$ impurity operators which we store in such a way to always preserve global 
time ordering along the contour. In summary a typical configuration reads
\be
\m{C}=
\left\{
\begin{array}{l}
a=1,2,\cdots,\m{N}\\ 
k_a=0,\ldots\infty\\
t_1^a \stackrel{C}{>}t_2^a\stackrel{C}{>}\ldots\stackrel{C}{>}t^a_{k_a}\\
\tilde{t}_1^a\stackrel{C}{>}\tilde{t}_2^a\stackrel{C}{>}\ldots\stackrel{C}{>}\tilde{t}^a_{k_a}\,.
\end{array}
\right.
\ee
An example of such a configuration is shown in figure~\ref{fig:config_example}.
The weight $W\left(\m{C}\right)$ associated to each configuration $\m{C}$ can be read directly from the hybridization expansion, 
see Eq. (\ref{eqn:hybrid_exp}). For later convenience we define it as
\be\label{eqn:weight}
W\left(\m{C}\right) = \mbox{Det}\left[\m{C}\right]\,\mbox{sign}\left[\m{C}\right]\,
\mbox{Tr}_{loc}\left[\m{C}\right]\,,
\ee
where $\mbox{sign}\left[\m{C}\right]$ includes all the signs(phases) coming from the evolution as well as from the time ordering, 
while the trace over the configuration reads
\be\label{eqn:loc_trace_2}
\mbox{Tr}_{loc}\left[\m{C}\right]=\mbox{Tr}\left[\rho_{loc}\,X_1\left(t_1\right)\,X_2\left(t_2\right)
\,\dots\,X_{2N}\left(t_{2N}\right)\,\right]\,,
\ee
so that by definition we get for $\m{Z}\left(t\right)$
\be\label{eqn:Z_diagMC}
\m{Z}\left(t\right)=\sum_{\m{C}}\,W\left(\m{C}\right)\,.
\ee
The same weights $W\left(\m{C}\right)$ are also required for evaluating the average of any local operator $O$ acting on the 
impurity Hilber space. Indeed, if consider the definition of quantum averages given in Eq.~(\ref{eqn:O_aver}), namely
\be\label{eqn:O_aver_2}
 \langle O(t)\rangle =
Tr\,\left[\rho_{eq}\,U^{\dagger}\left(t\right)\,\m{O}\,U\left(t\right)\right]\,,
\ee
and perform the hybridization expansion of the previous section, we find that 
\begin{equation}\label{eqn:O_aver_MC}
\langle O\left(t\right) \rangle = \frac{\sum_{\m{C}}\,\m{O}\left(\m{C}\right)W\left(\m{C}\right)}
{\sum_{\m{C}} W\left(\m{C}\right)}\,,
\end{equation}
where the estimator of local operator has been defined as
\be\label{eqn:estimator_local_ope}
O\left(\m{C}\right) = \frac{\mbox{Tr}\left[\rho_{loc}\,X_1\left(t_1\right)\,\,\dots\m{O}\dots\,X_{2N}\left(t_{2N}\right)\right]}
{\mbox{Tr}\left[\rho_{loc}\, X_1\left(t_1\right)\,\,\dots\,X_{2N}\left(t_{2N}\right)\right]}\,.
\ee

Once the real-time average of a local operator is written as in Eq. (\ref{eqn:O_aver_MC}), it would be natural
natural to sample it using a Monte Carlo method, namely generating a random walk in the configuration space which visit configurations $\m{C}$ with probability $P\left(\m{C}\right)=W\left(\m{C}\right)/\sum_{\m{C}'}W\left(\m{C}'\right)$.

When trying to implement this idea in the context of real-time quantum dynamics the problem one have to face is that the weight $W\left(\m{C}\right)$ is in general a complex number. In the specific case of interest this is not only due to the explicit ''$i$-factors'' coming from the real-time evoultion and entering $\mbox{sign}[C]$ but also to the fact that the contour ordered bath, defined in Eq (\ref{eqn:delta_contour}) and entering the determinants,  is indeed a complex function of its time arguments (see Appendix \ref{app:contour_ord_hyb}). The simplest way to circumvent this problem is to sample the absolute value of the weight, $\vert W\left(\m{C}\right) \vert$, while including the phase $\eta\left(\m{C}\right)$, defined as
\be\label{eqn:phase}
\eta\left(\m{C}\right)=\frac{W\left(\m{C}\right)}{\vert W\left(\m{C}\right)\vert}\,,
\ee
in the Monte Carlo estimator. While this approach allows for a straightforward implementation, it becomes problematic when the average phase goes to zero. In this respect, we note that more refined but computationally expensive techniques, based on sampling blocks of configurations at fixed sign, has been developed in recent years to cope with this dynamical sign problem\cite{Egger_Mulbacher}. Therefore a possible future direction could be to merge them with present implementations of diagMC method to see if a compromise between efficiency and accurancy can be found. For the time being we avoid this route, sampling directly the absolute value of the weight.

\subsection{Metropolis Algorithm}

The standard technique to generate configurations with a given probability (in the case of our interest $P_{\infty}\left(\m{C}\right)\equiv\vert W\left(\m{C}\right)\vert/\sum_{\m{C}'}\vert W\left(\m{C}'\right)\vert$) is to build up a Markov chain, namely a stochastic process which describes the evolution of the probability to visit configuration $\mathcal{C}$ after $n$ steps.
\be\label{eqn:proba}
P\left(\mathcal{C},n\right)= Proba\left(\mathcal{C}\left(n\right)=\mathcal{C}\right)
\ee
This Markov chain is fully determined once we assign the conditional probability   $\m{S}\left[\mathcal{C}'\vert\,\m{C}\right]$ to be in $\m{C}'$ at step $n+1$ being in $\m{C}$ at step $n$. Indeed, this is the quantity entering the master equation
\begin{equation}\label{eqn:master} 
P\left(\mathcal{C}',n+1\right) = \sum_{\m{C}}\;\m{S}\left[\m{C}'\vert\,\m{C}\right]\,P\left(\mathcal{C},n\right)\,.
\end{equation}
Sufficient conditions for this master equation to reach, after waiting a proper \emph{equilibration} time, the desired probability $P_{\infty}\left(\m{C}\right)$ is that the matrix $\m{S}\left[\m{C}'\vert\m{C}\right]$ is \emph{ergodic} and satisfies the so called detailed balance condition. Ergodicity means that it has to be possible to reach any configuration $\m{C}$ from any other configuration $\m{C}'$ in a finite number of steps, while detailed balance means that for any couple of configuration $\m{C}$ and $\m{C}'$ the following relation has to hold
\begin{equation}\label{eqn:detbal}
\m{S}\left[\m{C}'\vert\,\m{C}\right]P_{\infty}\left(\m{C}\right)=
\m{S}\left[\m{C}\vert\,\m{C}'\right]P_{\infty}\left(\m{C}'\right)\,,
\end{equation}
where $P_{\infty}$ is the probability distribution we want to sample through the Markov chain. 

A simple algorithm to generate configurations so to satisfy detailed balance was introduced by Metropolis\cite{Metropolis}. The idea is to start from a given intial configuration $\m{C}$ and to propose to visit a new configuration $\m{C}'$ with a certain transition probability $T\left(\m{C}'\vert\,\m{C}\right)$. Then this new configuration is accepted or rejected according to the probability $A\left(\m{C}'\vert\,\m{C}\right)$ so that the full conditional probability to move toward $\m{C}'$ starting from $\m{C}$ is given by
\be\label{eqn:prova_cond}
\m{S}\left[\m{C}'\vert\,\m{C}\right]=T\left(\m{C}'\vert\,\m{C}\right)\, 
A\left(\m{C}'\vert\,\m{C}\right)\,.
\ee
The Metropolis choice for the acceptance probability $\m{A}\left(\m{C}'\vert \m{C}\right)$ reads
\be\label{eqn:acceptance}
\m{A}\left(\m{C}'\vert \m{C}\right)=min\left\{1,\frac{P_{\infty}\left(\m{C}'\right)\,
T\left(\m{C}\vert \m{C}'\right)}{P_{\infty}\left(\m{C}\right)\,
T\left(\m{C}'\vert \m{C}\right)}\right\}\,.
\ee
It is easy to show that such a choice satisfies the detailed balance condition in Eq.~(\ref{eqn:detbal}).
While this algorithm is completely standard and model independent, two main issues have 
to be taken into account with reference to the problem at hand, since they can strongly affect the performance or even the reliability of the Monte Carlo simulations. 

The first one concerns the choice of the transition probability $T\left(\m{C}'\vert\,\m{C}\right)$. In the case of interest, we implement two main classes of local moves, in which the number of kinks in a given channel $a$ is changed by unity, $\Delta k_a=\pm1$. These moves amount to add or remove one creation and one annihilation fermionic operator in the $a$ channel at randomly chosen times along the contour and are required for the ergodicity of the matrix $\m{S}$. Indeed it is evident that, using these two basic updates any configuration can be reached, in principle, after a finite number of steps\footnote{One can imagine, given a configuration $\m{C}$ with $2k$ kinks, to remove all kinks and then add $2k$ new kinks at different times to obtain a new configuration $\m{C}'$. Wheter such a series of updates is likely to occurr during typical simulation time is a different matter. Additional updates may be useful to speed-up the process. This point is enligthned in ~\cite{Gull_PhD}.}. Therefore using these two classes of moves and the Metropolis acceptance Eq. (\ref{eqn:acceptance}) we can guarantee that the sampling process visits configurations according to probability $P_{\infty}\left(\m{C}\right)$. A second issue, which is different from the ergodicity one, concerns the efficiency and the speed-up of the Monte Carlo sampling (for example the number of steps which one has to wait before reaching the desired probability distribution). For this purpose, additional kind of updates may enhance the sampling procedure. 
In the present algorithm, following standard practice in diagMC, we also implement moves that connect configurations at fixed number of kinks ($\Delta k_a=0$) such as for example shifting an annihilation operator. We also note that other kind of moves, in which more than two operators are added/removed/shifted, may become relevant when dealing with off-diagonal baths. Similarly global moves, in which a whole set of operators is changed, has proven to be fundamental in the case of multiorbital models\cite{Michel_ctqmc}. We note that, as it happens for the imaginary time algorithm\cite{ctqmc_Werner}, a major issue in the implementation of these Monte Carlo moves is to properly take into account the structure of the impurity hilbert space to avoid, when it is possible, moves toward configurations which have zero weight. In the case of impurity models without exchange or hopping terms, these zero-weight configurations can be immediately read out since, for each channel $a$, creation and annihilation operators have to occur in alternated order along the contour, to have a finite local trace in Eq. (\ref{eqn:weight}). This leads to a very convenient segment picture\cite{ctqmc_Werner}.

A further point which requires some comment concerns the evaluation of the acceptance ratio in 
Eq.~(\ref{eqn:acceptance}). As can be seen from the expression of the weight $W\left(C\right)$ in 
Eq.~(\ref{eqn:weight}), this amounts to evaluate the ratio of two determinats as well as the ratio between two local traces. While for the former fast update routines are available, which makes this operation rather efficient, dealing with the ratio among local traces is the most time-consuming part of the algorithm, at least in the general case of a multi-orbital quantum impurity model. In such a case, indeed, one have to evaluate from scratch the whole chain of fermionic operators. Several tricks have been proposed to implement this evaluation\cite{Haule_ctqmc} in an efficient way and we have used most of them in our algorithm. In particular, we write the fermionic operators entering in Eq. (\ref{eqn:loc_trace_2}) in the basis of local eigenstates and store the whole chain of matrix products from left to right (and viceversa), so that the evaluation of trial moves is reduced to few matrix multiplications.

In the next section we describe the first application of the diagMC algorithm on the Kadanoff-Baym-Keldysh contour to the single impurity Anderson Model. In particular we will focus on the impurity real-time dynamics after a local quantum quench.

\section{Real-Time Dynamics in the Anderson Impurity Model after a local quantum quench}\label{sect:AIM}

Quantum quenches in strongly correlated systems have recently attracted lot of scientific interest,
especially inspired by exciting experiments on cold atomic gases\cite{Greiner_nature_02} where sudden changes of Hamiltonian parameters has been realized and the non equilibrium dynamics monitored in real-time.
In the context of impurity models, instead, the study of quantum quenches has a long history which goes back to the fundamental 
work by Nozi\`eres and De Dominicis on the X-ray edge singularity\cite{XrayEdge}, passing through the famous Anderson and Yuval 
approach to the Kondo model\cite{AY}. More recently, this problem stimulated new 
interest\cite{Nordlander_99,Kehrein_Lobaskin_05,tnrg_Anders}, due to the experimental progresses in nanotechnology, 
which made it possible to contact microscopic quantum objects with metallic electrodes, 
thus realizing quantum impurity models in a fully tunable set-up\cite{GoldhaberGordon_prl98}. Two kinds of quenches can be 
considered in this context, depending on the amount of energy that is injected into the system, also referred as the work done 
during the quench. Global quantum quenches are particularly relevant for transport through correlated nanostructures,  
where a net current flow is forced by suddenly switching on e.g. a dc bias voltage. 
Since the switched perturbation is extensive, the system is driven into a 
non-equilibrium steady state at long times\cite{Andrei_Doyon_PRB}. 
Conversely, local quantum quenches amount to suddenly change the impurity Hamiltonian. These kinds 
of quenches could be realized in an optical absorption experiment, as suggested in\cite{Glazman_quench}, and the 
resulting non-equilibrium dynamics can be tracked in real-time using pump-probe techniques or, in real-frequencies, 
measuring the absorption lineshape. Furthermore, local quenches are interesting as they are the simplest 
examples of non-equilibrium processes whose statistics may show non trivial fluctuations\cite{Silva_work_statistics}.

To test the algorithm, in this section we study the non-equilibrium dynamics in the Anderson Impurity Model\cite{AIM_meanfield} 
after a local quantum quench. This model, which serves as a fundamental paradigm for strong correlation physics, 
describes a single interacting fermionic orbital coupled to a an equilibrium bath of free conduction electrons.
The local hamiltonian before and after the quench reads
\be\label{eqn:local_hamilt}
\m{H}^{\pm}_{loc}\left[\cdag_{\sigma},c_{\sigma}\right]=
\frac{U_\pm}{2}\left(n-1\right)^2 + \eps_{d\pm}\,n\,
\qquad n=\sum_{\sigma}\,c^\dagger_{\sigma}c_{\sigma}\,.
\ee
The conduction electrons are assumed to be non interacting, hence the coupling to the impurity occurs via  
an energy-dependent hybridization function $\Gamma\left(\eps\right)$, which is defined in terms of 
the conduction density of states (DoS) 
$\rho\left(\eps\right)$ as
\be\label{eqn:hybrid}
\Gamma(\eps)=\pi\,\sum_{\mathbf{k}}\,\vert V_{\mathbf{k}}\vert^2\,\delta(\eps-\eps_\mathbf{k})=\pi\,V^2\,\rho\left(\eps\right)\,,
\ee
where we have assumed for simplicity $V_{\mathbf{k}}$ independent of momentum. As a model for the DoS we start considering a flat band of width $2W$
\be\label{eqn:flat_dos}
\rho\left(\eps\right)=\rho_0\,\theta\left(W-\vert\eps\vert\right)\,,
\ee
which encodes the main physics of a metallic conduction bath, namely a finite weight at the Fermi level ($\rho_0=1/2W$ at $\eps=0$) and a finite bandwidth. In this case the basic energy scale describing the coupling between the impurity and the bath is the hybridization strength $\Gamma=\pi V^2\rho_0$. In all calculations we take $\Gamma$ as our unit of energy and choose $W=10\Gamma$.

This section is structured as follows. We first discuss some aspect of the algorithm (statistics of kinks and average sign) in the specific case of the  Anderson Impurity Model and then present the results for its charge and spin real-time dynamics after a local quantum quench.

\subsection{Performance Analysis of the Kadanoff-Baym-Keldysh diagMC algorithm}

In order to analize the performances of the diagMC algorithm on the Kadanoff-Baym-Keldysh contour we will consider two main quantities, namely the probability distribution of perturbative orders in the diagrammatic expansion and the average sign of the Monte Carlo weight, both being sensitive measures to establish the efficiency of the method. In the specific case of the single impurity Anderson Model, with reference to the notation introduced in section \ref{sect:hybrid_exp}, we have only $\m{N}=2$ channels, corresponding to spin $\sigma=\up,\dw$.

\subsubsection{Statistics of Kinks}

\begin{figure}[t]
\begin{center}
\epsfig{figure=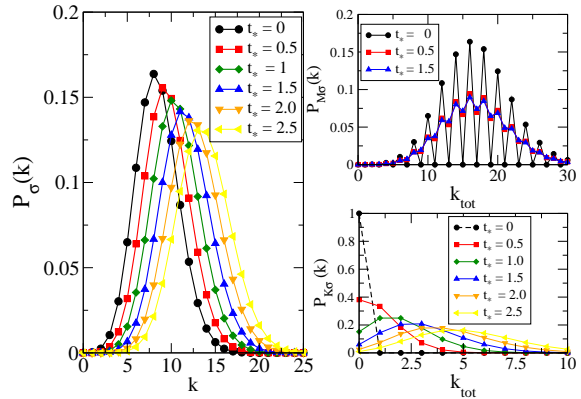,scale=0.3}
\caption{Probability distribution of different perturbative orders $k$ sampled during the simulation. Data refers to a local quench in the Anderson Impurity Model, starting from $U_-=0$ to $U_+=10\Gamma$ at particle-hole symmetry, for $T=0.1\Gamma$ and $W=10\Gamma$. Left panel shows the statistics ok kinks along the whole contour, while the top and bottom right panels display the hystograms for the Matsubara and Keldysh sector, respectively.}
\label{fig:hysto_vs_tmax_Uin_0_Ufin_0} 
\end{center}
\end{figure}

As we have shown in section~\ref{sect:diagMC}, diagMC amounts to stochastically sample the expansion for 
$\m{Z}\left(t\right)$ in powers of the hybridization, by performing a random walk in the space of diagrams. 
It is therefore quite natural to monitor during the simulation the statistics of different perturbative orders (number of kinks), 
namely the probability to visit a Monte Carlo configuration $\m{C}$ featuring $k$ creation vertices (and $\tilde{k}$ annihilation 
vertices) in the spin sector $\sigma$. We define this quantity as
\be\label{eqn:proba_k}
P_{\sigma}\left(k\right)=\frac{\sum_{\m{C}}\,\vert W\left(\m{C}\right)\vert\,\delta\left(k\left(\m{C}_{\sigma}\right)-k\right)}
{\sum_{\m{C}}\,\vert W\left(\m{C}\right)\vert}\,,
\ee
where the Monte Carlo weight $W\left(\m{C}\right)$ has been defined in Eq.~(\ref{eqn:weight}). 
The behaviour of this probability distribution is plotted, as an example, in the left panel of 
figure~\ref{fig:hysto_vs_tmax_Uin_0_Ufin_0} for increasing values of the measuring time $t_{\star}$ starting from
$t_{\star}=0$, which corresponds to the equilibrium initial preparation. 
We note that all hystograms are strongly peaked around an average value $\bar{k}$, larger perturbative orders $k>\bar{k}$ appearing with an exponentially small probability. Notice that whenever this would not be the case, namely if arbitrarily large perturbative orders ($k\gg\bar{k}$) would give a finite contribution to the result, a diagMC algorithm to work would need an explicit cut-off $k_{max}$ on the perturbative order and the result would then require an extrapolation\footnote{This is what happens, for example, in diagMC simulation of lattice models with the bold-line trick\cite{Prokofev_sign}, where the fermionic sign problem is so severe that a simulation with arbitrary (unbounded) self-consistent perturbative orders turns to be unstable and an explicit cut-off is needed.}  to $k_{max}\rightarrow\infty$.  However for quantum impurity models, at least for the weak and strong coupling algorithms\cite{ctqmc_Rubtsov,ctqmc_Werner,Haule_ctqmc}, this is not the case. Figure~\ref{fig:hysto_vs_tmax_Uin_0_Ufin_0} confirms that all orders are included and contribute, with their own weight, to the final result. This fact ensures that the outcome of our diagMC calculation is an unbiased result which do not correspond to any truncation at finite-order of the perturbative expansion but rather represents a numerical resummation of a formal expansion. 
From figure~\ref{fig:hysto_vs_tmax_Uin_0_Ufin_0} we note that upon increasing $t_{\star}$ the whole hystogram shifts toward 
larger values of $k$ since kinks start to be added on the two Keldysh branhces. It is therefore interesting to \emph{resolve} 
this increased perturbative order in the two sectors of the simulation, namely the Matsubara and the Keldysh one. 
To this extent, we plot in the right panel of the same figure the probability distribution of having $k_{tot}$ kinks with spin $\sigma$ 
on the Matsubara axis, $P_{M\sigma}\left(k\right)$ (top right panel), or $k_{tot}$ kinks with spin $\sigma$ on the Keldysh branches, 
$P_{K\sigma}\left(k\right)$ (bottom right panel), where $k_{tot}$ means that we are considering both creation and annihilation vertex.
At $t_{\star}=0$, the initial state, the Keldysh branches are empty while the 
Matsubara sector is filled with an even number of kinks (to ensure particle  conservation). Upon increasing $t_{\star}>0$, 
the system starts evolving in real-time and we note a transfer of weight in the Keldysh sector toward finite values of $k$. 
At the same time the Matsubara probability distribution does not change its center of gravity and rapidly converges to a 
final distribution, now allowing also for an odd number of kinks (the total particle conservation is ensured by kinks in the Keldysh sector).
The scaling of the average number of kinks $\bar{k}$ with measuring time $t_{\star}$, temperature and other physical parameters 
is also relevant and instructive. In the equilibrium case (corresponding here to $t_{\star}=0$), $\bar{k}$ has been shown in 
Ref.~\cite{Haule_ctqmc} to be proportional to inverse temperature $\beta$ with a prefactor given by the average 
hybridization energy per spin,
\be\label{eqn:kaver_eq}
\bar{k}_{\sigma}=-\beta\langle\m{H}^{\sigma}_{hyb}\rangle.
\ee
Since $\langle\m{H}_{hyb}\rangle$ decreases upon increasing the correlation strength $U$ the diagMC method in imaginary time works 
extremely well in the regime $U\gg\Gamma$ being able to reach very low temperatures compared to the energy scales in the problem. 
\begin{figure}[t]
\begin{center}
\epsfig{figure=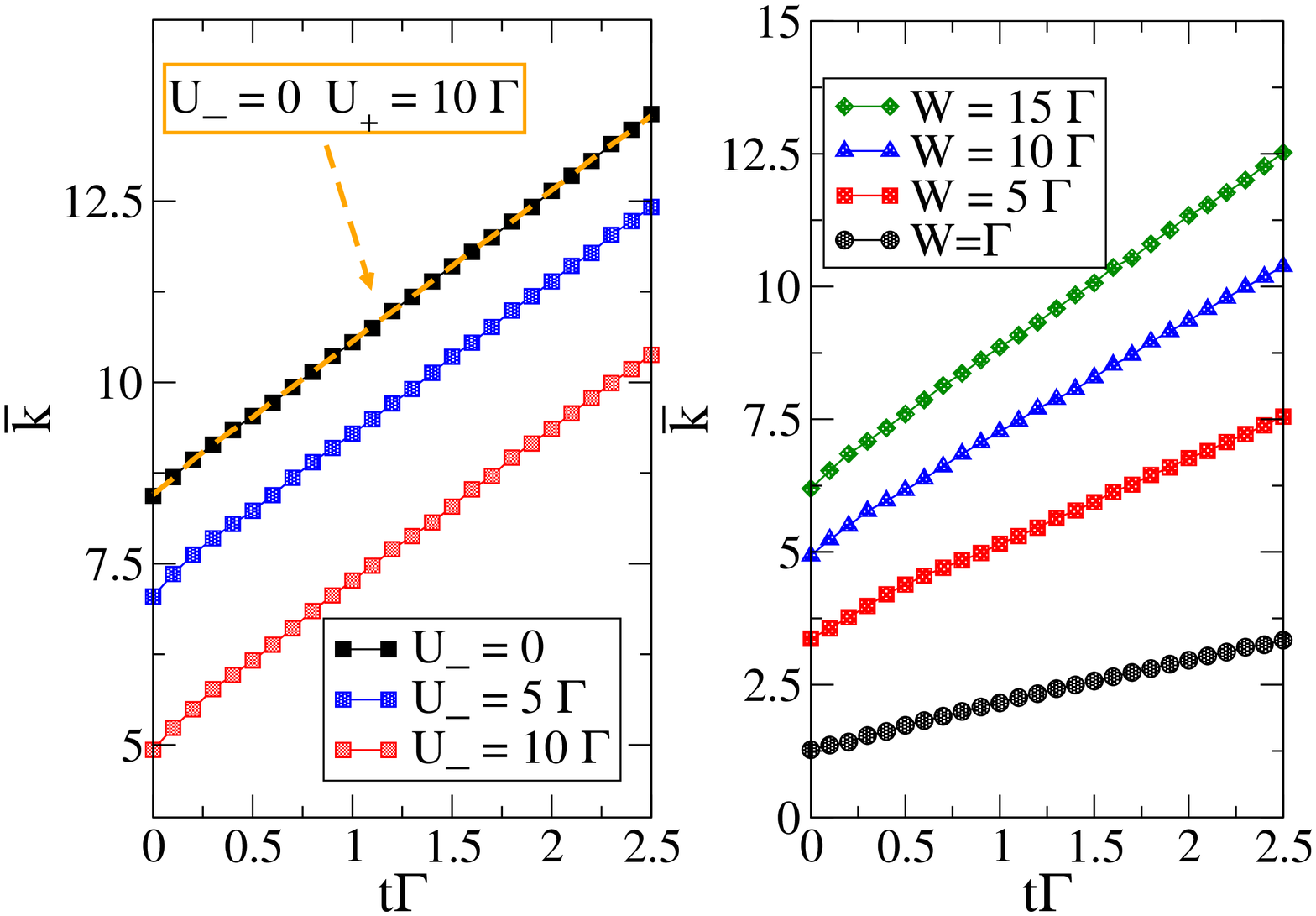,scale=0.3}
\caption{Scaling of the average number of kinks with maximum time $t_{\star}$. Left panel shows data at fixed $U_+=0$, tuning the strength of the Coulomb repulsion $U_-=0,5,10$. See the perfect linear scaling with the same slope $\alpha=d\bar{k}/dt$. Right panel displays data at fixed $U_-=10$, $U_+=0$ tuning the strenght of the conduction bandwidth $W$. We see how the slope $\alpha$ increases with bandwidth making increasingly difficult to access large time scales in the regime $W\gg\Gamma$.}
%
\label{fig:k_aver_vs_tmax} 
\end{center}
\end{figure}
Unfortunately, the very convenient scaling of Eq.~(\ref{eqn:kaver_eq}) does not hold anymore for the real-time dynamics, 
as was also noted in previous works\cite{Werner_Keldysh_09}. In figure~\ref{fig:k_aver_vs_tmax} (left panel) we plot 
$\bar{k}$ as a function of time $t_{\star}$ for different initial preparations $U_-=0,5,10$. We note an almost perfect linear 
scaling with time, as expected, while the effect of starting from a correlated initial state $U_-\neq 0$ generally helps since it 
decrease the value of $\bar{k}$ at $t=0$. 
\begin{figure}[t]
\begin{center}
\epsfig{figure=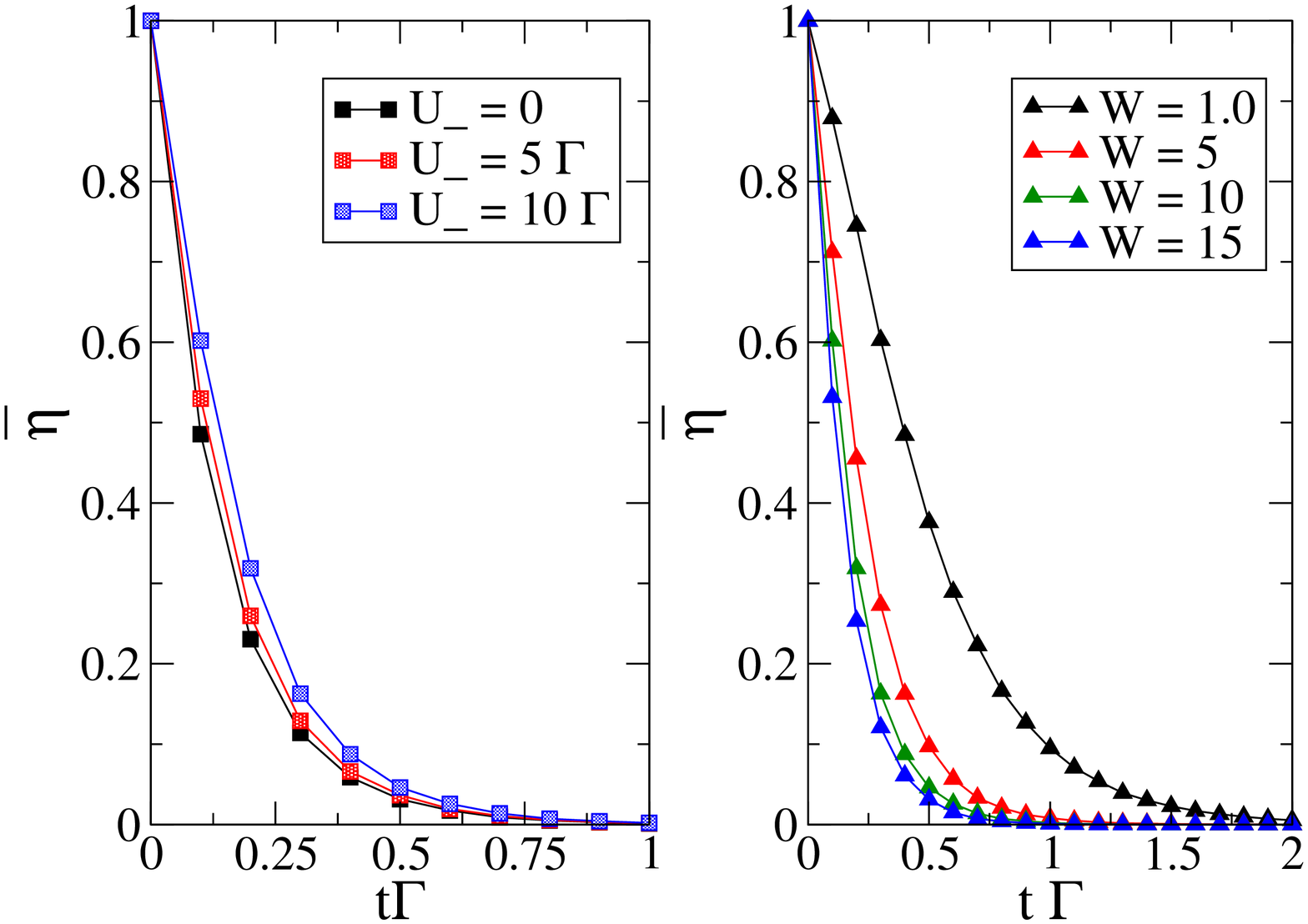,scale=0.3}
\caption{Average Sign as a function of time $t$ for different initial preparations. We clearly see an exponential decay on a vey short-time scale. Left panel shows data obtained fixing the final value of the interaction $U_+=0$ and tuning the initial value $U_=0,5,10$. We see a slight increase of the average sign. Right panel shows the dependence of $\bar{\eta}$ from the bandwidth of conduction electrons and suggest that much longer time scales can be reached in the regime $W\sim\Gamma$.}
\label{fig:sign_aver} 
\end{center}
\end{figure}
Nonetheless, a finite Coulomb interaction in the final state, 
$U_+\neq0$, has no effect on the average number of kinks sampled, as shown by the dashed line in left panel which exactly 
lies on top of the $U_+=0$ results. 

Summarizing, we conclude that the scaling of the average number of diagrams for the real-time algorithm generally reads
\be\label{eqn:kaver_noneq}
\bar{k}_{\sigma}=\alpha\,t + \bar{k}^{eq}_{\sigma}\,,
\ee
$\alpha$ being a costant independent on $U$. It is therefore natural to ask what is the energy scale controlling this prefactor. 
As we show in the right panel of Figure \ref{fig:k_aver_vs_tmax}, $\alpha$ strongly increases with the conduction 
bandwith $W$ (and presumably also on the hybridization strength). 
As a consequence of Eq.~(\ref{eqn:kaver_noneq}), accessing large time scales in the regime $W\gg\Gamma$ becomes increasingly 
difficult with this approach. This is due to the fact that both the computational cost of the algorithm and, in particular, 
the average sign of the MC weights strongly depend on the average number of kinks $\bar{k}$, exponentially the former and power-law the latter.

\subsubsection{Average Sign}

Another important quantity to monitor during the simulation is the average sign of the MonteCarlo configurations, 
which is tightly related to the accuracy we can get on physical quantities at fixed CPU time.
Indeed a vanishing average sign turns into very large error bars on MonteCarlo averages that makes the simulation unstable. 
In the specific case of the hybridization expansion diagMC, it is known that, for what concerns the equilibrium (imaginary-time) 
algorithm, the single impurity Anderson Model has always positive signs and that even multi-orbital impurity models 
with rotationally-invariant interaction can be efficiently simulated up to moderate low temperatures. This situation drastically 
change when dealing with real-time dynamics since even the simple non-interacting resonant level model faces a severe sign 
problem at large enough times scales. This seems to be due to the intrinsinc nature of unitary quantum evolution and clearly appears 
from the definition we gave of the dynamical partition function $\m{Z}\left(t\right)$ in section \ref{sect:hybrid_exp}. Indeed exact cancellations are built in the whole formalism to ensure unitarity.

\begin{figure}[t]
\begin{center}
\epsfig{figure=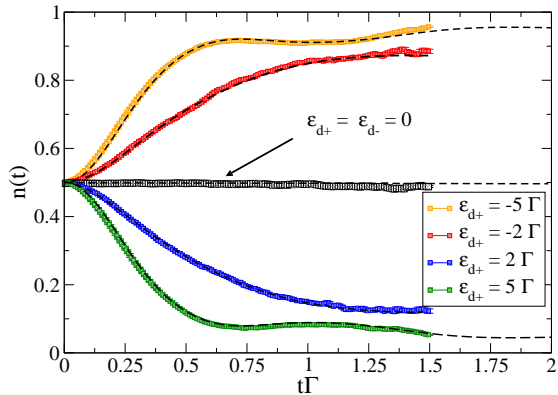,scale=0.3}
\caption{Quench dynamics in a Resonant Level Model after a sudden change of the energy level from $\eps_{d-}=0$ to $\eps_{d+}\neq0$. Dashed lines is the exact solution for $n\left(t\right)$ as obtained by a standard methods. Points are diagMC results obtained at $T=0.1\Gamma$. We also add the dynamics for the trivial case $\eps_{d+}=\eps_{d-}=0$ to show that unitarity is actually preserved by diagMC.}
\label{fig:benchmark_rlm}
\end{center}
\end{figure}
In the real-time diagMC one should in general talk about average phase, since as we mentioned the Monte Carlo weights are complex numbers. Nevertheless this quantity, which is defined as
\be\label{eqn:aver_MC_sign}
\bar{\eta}\left(t\right) = \frac{\sum_{\m{C}}\,\vert W\left(\m{C}\right)\vert\,\eta\left(\m{C}\right)}{\sum_{\m{C}}\,\vert W\left(\m{C}\right)\vert}\,.
\ee
turns to be directly related to the probability of visiting configurations in the Matsubara sector, namely to the probability of having no kinks on the real-time branches. As a consequence of this result, which comes directly from unitarity, we conclude the the average phase is indeed a purely real number even for $t>0$ and therefore, without further misunderstanding, we refer to it as the average Monte Carlo sign.
As we show in figure \ref{fig:sign_aver}, $\bar{\eta}\left(t\right)$ depends exponentially on the measuring time $t_{\star}$, namely on the length of the Keldysh contour. In the left panel, we plot the average sign for different values of the Coulomb repulsion $U_{-}$ in the initial density matrix. We see that $\bar{\eta}\left(t\right)$ uniformly increases with $U_{-}$, namely starting from a correlated initial state may result into a larger average sign even if the effect is rather small. In the left panel we study how the sign depends on the bandwidth $W$ of conduction electrons. We see that moving from $W=\Gamma$ to $W=15\Gamma$ there is a sizeable increasing of the average sign, which means that larger time scales can be reached with present algorithm in the regime $W\simeq\Gamma$. While this is not of direct relevance for quantum impurities, it can be very intersting for Non Equilibrium DMFT, where the conduction bandwidth is of the same order as the coupling between the impurity and the bath itself. 

\begin{figure}[t]
\begin{center}
\epsfig{figure=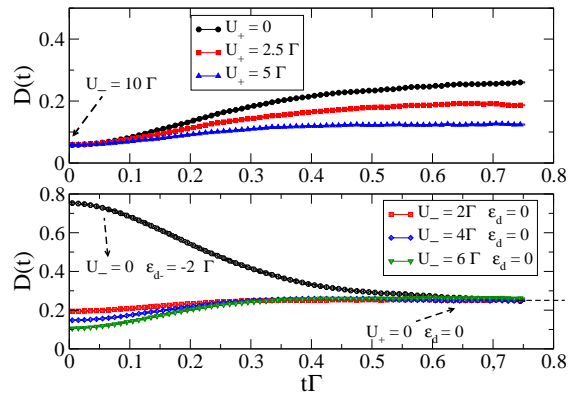,scale=0.3}
\caption{Non Equilibrium Dynamics of double occupancy $D\left(t\right)$ in the Anderson Impurity Model after a local quantum quench of the interaction strength at $T=0.1\Gamma$ and particle-hole symmetry. In the upper panel we start from an initial state with $U_-=10\Gamma$ and a very low double occupation and quench to different values of $U_+/\Gamma=0,2.5,5$ from top to bottom. In the lower panel the opposite protocol is considered, namely we start from different values of $U_-/\Gamma=2,4,6$ from top to bottom and quench to the same final $U_+$=0. In both cases we see that after a rather short transient the system relaxes to a new equilibrium state.}
\label{fig:double_occ_dyn_metallic} 
\end{center}
\end{figure}

\subsection{Charge and Spin Dynamics in the Anderson Model after a local quantum quench}\label{sec:charge_spin}

We start by considering the non interacting case, the so called resonant level model with $U_-=U_+=0$, which allows for an exact solution and can be therefore used to benchmark the algorithm. We consider for this simple resonant level model a quench of the energy level $\eps_d$ that we tune from 
the on-resonance value $\eps_-=0$ to the off-resonance one $\eps_+\neq0$. We note that this kind of quench can be realized in 
optical absorption experiments with quantum dots, as recently proposed in Ref.~\onlinecite{Glazman_quench}. In 
Figure~\ref{fig:benchmark_rlm} we plot the real-time dynamics of the impurity density $n\left(t\right)$ for two different quenches, 
respectively above and below the on-resonance value $\eps_d=0$, and compare the result of diagMC (datapoints) with the exact 
dynamics which can be obtained using standard methods\cite{Mahan_book}. The excellent agreement with exact results confirms the 
reliability of our numerical approach.

We then move to the interacting case, namely consider a local quantum quench in the Anderson Model with local 
Hamiltonian (\ref{eqn:local_hamilt}). In Figure~\ref{fig:double_occ_dyn_metallic}  we show the dynamics of the 
double occupation $D\left(t\right)=\langle n_\up\left(t\right)\,n_\dw\left(t\right)\rangle$ after a sudden quench of the local
interaction strength $U$. Two different cases are considered. In the upper panel of Figure~\ref{fig:double_occ_dyn_metallic},   
we start from the same initial preparation, $U_-=10\Gamma$, and quench to different final values of the interaction 
$U_+/\Gamma=0,2.5,5$ (from top to bottom). In the lower panel of the same figure, we start from different initial preparations 
$U_-/\Gamma=2,4,6$ (from top to bottom) and quench to the same final state $U_+=0$. 

\begin{figure}[t]
\begin{center}
\epsfig{figure=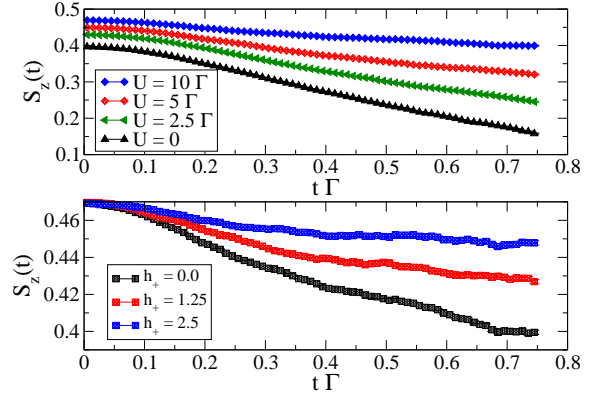,scale=0.3}
\caption{Spin Dynamics in the Anderson Model after a sudden quench of a local magnetic field. We start from a partially polarized impurity, with $h_-=5.0$, $T=0.1\Gamma$ and different values of local interaction $U$ (top panel). At time $t>0$ the magnetic field is switched off and the magnetization is allowed to relax toward an unpolarized steady state. We see that upon increasing the interaction $U$ the dynamics slows down. Due to fine-time resolution we cannot follow the decay of the spin toward zero magnetization.  However a large magnetic field in the final state gives rise to a rather fast relaxation.}
\label{fig:spin_dynamics} 
\end{center}
\end{figure}

The dynamics at short times, soon after the quench, is controlled by the initial density matrix as expected on general grounds. 
After a short time scale, $t_{short}\sim 0.1/\Gamma$, the system starts feeling the quench and in fact  
the curves in the upper/lower panel start to deviate from/approach to each other. The time scale controlling the approach to the steady state is set mainly by $1/\Gamma$ - without coupling to the bath no dynamics for the charge would arise at all. However the final value of the interaction also affects the dynamics, as one can see from data in the top panel of figure~\ref{fig:double_occ_dyn_metallic}.
We also compare these findings with the non-interacting case, where the quench is performed on the energy level which is suddenly placed out of resonance (see lower panel black curve). 
\begin{figure}[t]
\begin{center}
\epsfig{figure=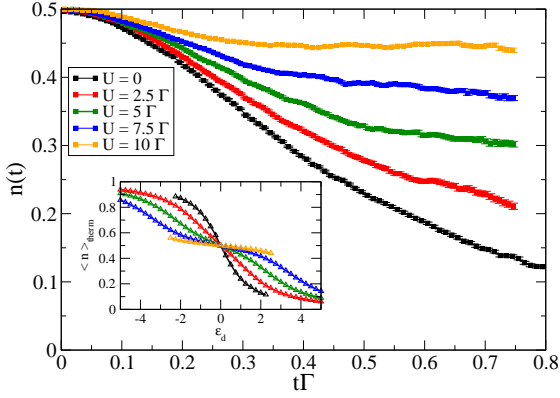,scale=0.3}
\caption{Non equilibrium dynamics of the impurity density after a quench of the energy level from $\eps_{d-}=0$ to $\eps_{d+}=2.5\Gamma$. We compare the dynamics for different values of the Coulomb repulsion $U=0,2.5,5,7.5,10$ from bottom to top, revealing a much faster thermalization in the correlated case due to Coulomb blockade. Inset: thermal value of the impurity density as a function of the level position for $U=0,2.5,5,7.5,10$ from top to bottom. For $U\gg\Gamma$ the curve is almost flat around $\eps_d=0$ hence departure from equilibrium is suppressed after a short transient.}
\label{fig:quench_gate_metallic} 
\end{center}
\end{figure}
In this situation the dynamics appears much slower than the previous cases, at least a factor of two. 
In Figure~\ref{fig:quench_gate_metallic} the problem of quenching the impurity energy level is considered for different values 
of the Coulomb repulsion $U$, starting from a level which is initially half-filled. As compared to the non interacting $U=0$ case, 
the effect of interaction is to make the whole relaxation dynamics much faster and the steady state value closer to the starting one, 
resulting in some sense into a less pronounced deviation from equilibrium. This can be rationalized by considering how the density 
depends, in equilibrium, on the energy level (see inset): upon increasing the interaction the curve $n\left(\eps_d\right)$ 
becomes flat around $\eps_d=0$, a signature of Coulomb blockade phenomenon. As a consequence, any perturbation which moves  
the impurity occupation out of integer filling is quickly suppressed on a short-time scale. The overall picture confirms 
what also found in a similar investigation with the time-dependent NRG in \onlinecite{tnrg_Anders}, namely that charge dynamics 
is sensitive to high-energy scales, thus resulting into a generally fast relaxation. As opposite to the charge sector, 
the dynamics of spin degree of freedom is sensitive to the low-energy physics of the model. To probe this dynamics in 
real-time we imagine to add a magnetic field $h_-$ to the local Hamiltonian, which partially polarizes the impurity, 
and suddenly switch it off for $t>0$. The local Hamiltonian Eq. (\ref{eqn:local_hamilt}) now reads
\be\label{eqn:local_hamilt_zeeman}
\m{H}^{\pm}_{loc}\left(h_{\pm}\right)=
\m{H}^{\pm}_{loc}\left(h_{\pm}=0\right) -\frac{h_{\pm}}{2}
\sum_{\sigma}\,\sigma n_{\sigma}
\ee
In Figure \ref{fig:spin_dynamics} we plot the spin dynamics $\langle S_z\left(t\right)\rangle$ starting from 
$h_-=5\Gamma$ and switching it off, $h_+=0$, for different values of the interaction $U$. 
Since the final state in the absence of Zeeman splitting is fully simmetric, we expect to recover, for large enough times, 
a relaxation to an unpolarized state with $\langle S_z\rangle=0$. We see that this relaxation is very slow and controlled 
by a time scale which \emph{increases} with $U$ (see top panel), as opposite to what found in previous cases, 
when the dynamics of charge degrees of freedom was probed by quenching the interaction or the level position. 
Indeed, the spin dynamics in the strong coupling regime is controlled by the lowest energy scale in the problem, 
namely the Kondo temperature, as explicitly shown in \cite{tnrg_Anders}. Accessing such a long time scale seems 
so far unfeasable within the present approach, since diagMC simulations become increasingly unaccurate at large times due to 
sign problem, as we will discuss in the next section.
As shown in the bottom panel of Figure \ref{fig:spin_dynamics}, the effect of a large magnetic field in the final state is to destroy Kondo effect, thus resulting again into a fast relaxation toward a new steady state. An interesting direction for future investigation is to study the spin dynamics in the Kondo regime under the effect of more general non equilibrium processes, other than a quantum quenches. In this respect the present approach can deal with explicitly time dependent phenomena, such as for example an oscillating magnetic field,  without any truncation of the dynamics.

\subsection{Non Equilibrium Dynamics for a quantum impurity in a gapped or pseudogapped fermionic reservoir}

In all cases considered above, we observe at large times a convergence to a 
new equilibrium state which is the thermal one described by the final hamiltonian $\m{H}_+$, namely
\be\label{eqn:thermalization}
\langle O\left(t\rightarrow\infty\right)\rangle=
\frac{\mbox{Tr}\left[e^{-\beta\m{H}_+}\,O\right]}
{\mbox{Tr}\left[e^{-\beta\m{H}_+}\right]}
\ee
This is explicitly shown by the dashed line in figure~\ref{fig:double_occ_dyn_metallic}, which represent the result of an 
equilibrium calculation done with imaginary time diagMC with the final Hamiltonian $\m{H}_+$. 
We also note that no effective heating arises, namely the temperature entering in Eq.~(\ref{eqn:thermalization}) is the same 
as in the initial condition (\ref{eqn:rho_therm}). 
This is due to the fact that within diagMC the fermionic reservoir is treated as an infinite system. 
The onset of thermalization in a  quantum impurity model is not surprising\cite{Caldeira_Leggett_PRL81}, and it is related to 
the fact that the conduction electrons play the role of a thermal bath\cite{Andrei_Doyon_PRB}, able to 
absorbe the energy pumped locally after quench, which is dissolved in the interior of the bulk.  
It is worth noting that this feature is not generic of any bath -- meant as a macroscopic (infinite) system -- but rather depends on its spectral properties. In the present case, as we are going to see, thermalization is related to the \emph{gapless} nature of the metallic state, whose energy spectrum goes down to arbitrarily small energies. To further investigate this issue, we consider now the out-of-equilibrium dynamics of an Anderson impurity coupled to gapped and pseudo-gapped fermionic bath. Even though QI models traditionally deal with genuine metallic hosts, the problem of gapped (or pseudo-gapped) fermionic reservoirs has a vast literature\cite{Fradkin_PRL,Ingersent_PRB,Cornaglia_PRL,CastroNeto}, that received a large boost in recent years. Eminent examples of such a physical situation are provided by adatoms in graphene sheet or by nanostructures built up with superconducting materials.

The equilibrium phase diagram of an Anderson impurity embedded in a non-metallic host it is by now rather well established\cite{Ingersent_PRB,Chen_PRB,Pruschke_RMP}. As opposite, the non equilibrium real-time dynamics in this class of quantum impurity models is much less explored. A detailed study of this issue goes well beyond the scope of this paper and it will be left for future investigations. Here we limit to elucidate the role played by the presence/absence of low energy bath spectral weight on the single impurity dynamics after a local quantum quench. It is worth to notice that this issue can be also relevant to study, within Non Equilibrium DMFT, the relaxation dynamics of interacting electrons after quantum quenches. 
Indeed DMFT amounts to solve a quantum impurity self-consistently, using the contour ordered impurity Green's function as a seed to generate the new fermionic out-of-equilibrum bath.

\subsubsection{Gapped Fermionic Reservoir}

\begin{figure}[t]
\begin{center}
\epsfig{figure=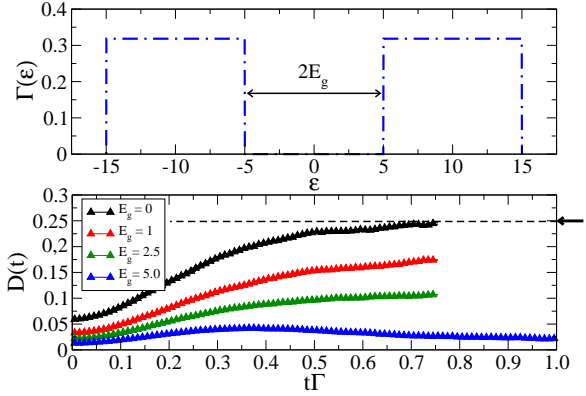,scale=0.3}
\caption{Non equilibrium dynamics for an Anderson Impurity coupled to a \emph{gapped} fermionic reservoir. We plot the real-time dynamics for double occupancy at the impurity site after a quench of the interaction from $U_-=10\Gamma$ to $U_+=0$, at particle-hole symmetry $\eps_{d+}=\eps_{d-}=0$ and for $T=0.1\Gamma$. Different values of the gap in the bath $E_g=0,1,2.5,5.0$ are considered, resulting into very different dynamics. Contrarily to the gapless case ($E_g=0$, black points) which quickly approaches the thermal plateau fixed by PH symmetry and indicated by an arrow, $D_{therm}=1/4$,  we see that due to the finite gap in the spectrum the real-time dynamics slows down thus preventing us to conclude on the long time behaviour of $D(t)$. However, for very large values of $E_g$ (see $E_g=5\Gamma$) the dynamics seems actually to reach a steady state where the double occupation is different from $D_{therm}$.}
\label{fig:fix_quench_tune_gap} 
\end{center}
\end{figure}

We start our discussion considering the case of a true gapped fermionic bath. In other words we consider as a model for the conduction electrons DoS the following
\be\label{eqn:band_gap}
\rho_g\left(\eps\right)=
\left\{
\begin{array}{lll}
0 & &0<\vert\eps\vert< E_g \\
\rho_0& &E_g<\vert\eps\vert<E_g+W 
\end{array}
\right.\,,
\ee
where $2E_g$ is the band gap at the Fermi Level. This density of states results into an energy dependent hybridization function $\Gamma\left(\eps\right)$ that we define as in Eq. (\ref{eqn:hybrid}), namely $\Gamma\left(\eps\right)=\pi\,V^2\rho\left(\eps\right)$. A plot of this function for different values of $E_g$ is given in figure~\ref{fig:fix_quench_tune_gap}. We note that in the following we will adopt as unit of energy the hybridization width $\Gamma=\pi\,V^2/2W$, the same as in the metallic case.

The equilibrium properties of an Anderson Impurity coupled to a gapped reservoir have been studied with NRG in \cite{Chen_PRB} and more recently with a perturbative approach in~\cite{Logan_PRB}. The model at particle-hole (PH) symmetry flows at low temperature to the local moment fixed point where the impurity is asymptotically decoupled from the bath. Out of PH the model displays a transition between local moment fixed  point and strong coupling fixed point depending on wheter the gap $E_g$ is larger or smaller than the Kondo temperature.  Here we consider for simplicity the PH symmetric point which correspond to setting $\eps_{d-}=\eps_{d+}=0$ in the local hamiltonian Eq. (\ref{eqn:local_hamilt}) and discuss the real-time dynamics for the double occupancy $D(t)=\langle n_\up\left(t\right)\,n_\dw\left(t\right)\rangle$ on the impurity site after a sudden change of the local Coulomb interaction. 

In figure~\ref{fig:fix_quench_tune_gap} we show the double occupancy dynamics after a quench from $U_-=10\Gamma$ to $U_+=0$ for different values of $E_g$ at $T=0.1\Gamma$. Due to PH symmetry the thermal value of $D$ computed on the final hamiltonian $\m{H}_+$ has to be equal to $D_{therm}=1/4$ for $U_+=0$. Indeed, we see that in the metallic case ($E_g=0$) $D(t)$ approaches rather quickly the expected thermal plateau. At the same time opening a finite gap $E_g\neq0$ in the bath reflects into a much slower dynamics which prevents us from a firm conclusion on the asymptotic behaviour of $D(t)$. We note, however, that for large values of $E_g$ the dynamics seems actually to reach a stationary state which looks quite different from the expected one.
Such a behaviour could be interpreted in terms of the equilibrium properties of the gapped Anderson impurity model which, as we mentioned, at PH symmetry flows to the local moment regime with the impurity effectively decoupled from the bath. 
\begin{figure}[h]
\begin{center}
\epsfig{figure=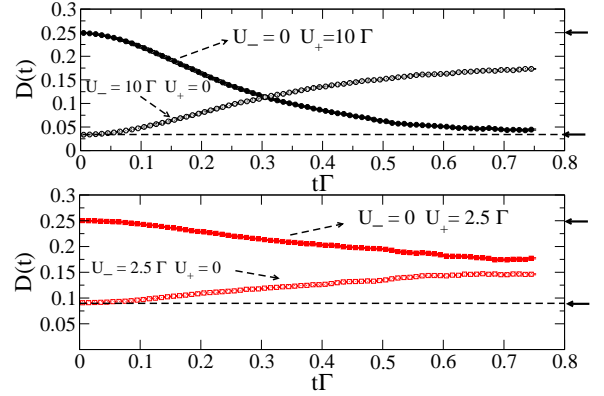,scale=0.3}
\caption{Non equilibrium dynamics for an Anderson Impurity coupled to a \emph{gapped} fermionic reservoir. We set the gap in the spectrum equal to $E_g=1.0$ and plot the real-time dynamics for the double occupancy after a quench of the interaction, at $T=0.1\Gamma$ and PH symmetry. Two kind of processes are considered, namely a quench from $U_-=0$ to $U_+\neq0$ and the reverse process from $U_-\neq0$ to $U_+=0$, which differ among each other for the average work done during the quench, see Eq. (\ref{eqn:work_local_U}) in the main text. Top panel shows data for $U_-=10\Gamma$, $U_+=0$ and viceversa, while bottom panel data for $U_-=2.5\Gamma$, $U_+=0$ and viceversa. We see that, provided the average work done is sufficiently larger than the gap $2E_g$, a fast thermalization can occurr also in a gapped model (see top panel, black points). As opposite, when the amount of energy pumped into the system is too small the dynamics slows down and we cannot conclude with present data wheter thermalization takes place or not.}
\label{fig:fix_gap_tune_quench} 
\end{center}
\end{figure}
Given such an initial condition and taking into account the large value of the gap,
which strongly affects the bath properties, one can rationalize the slowing-down of the impurity dynamics. Indeed, in the limit of very large gaps $E_g\rightarrow\infty$, a free impurity would have no available mechanism to exchange energy and relax to the steady state described by $\m{H}_+$. While this argument could be in principle satisfying to explain results plotted in figure~\ref{fig:fix_quench_tune_gap}, at least in the large gap regime, it does not take into account completely the nature of a quantum quench process. To further investigate this point we now reverse the perspective, namely we fix the gap $E_g$ in the spectrum and change the strength of the quench, namely we change the final value of the interaction $U_+$ while keeping fixed $U_-=0$ as well as the level position so to ensure PH symmetry. This allows to study how the non equilibrium dynamics depends on the amount of \emph{work done} during the quantum quench. As was recently suggested in\cite{Silva_work_statistics} the statistics of the work is a key quantity to characterize a non equilibrium process such as a quantum quench. Its average value $\bar{W}$ gives a measure of the energy pumped into the system and turns to be given\cite{Silva_work_statistics}, in the case of an istantaneous quench, by
\be\label{eqn:average_work}
\bar{W} = \langle\m{H}_- - \m{H}_+\rangle_{-}\,,
\ee
where the average $\langle\,\cdot\,\rangle_{-}$ is taken over the initial equilibrium density matrix $\rho_{eq}\propto e^{-\beta\m{H}_-}$. In the case of a local quantum quench such as the one we are considering, the average work $\bar{W}$ is given by
\be\label{eqn:work_local_U}
\bar{W} = (U_--U_+)\langle D\rangle_{-}\,.
\ee
We see therefore that the work $\bar{W}$ depends not only on the strength of the quench, namely the change in the interaction, but also on the initial condition. As we are going to see, this quantity greatly affects the resulting non equilibrium dynamics.

In figure~\ref{fig:fix_gap_tune_quench} (top panel) we plot the dynamics of double occupancy $D(t)$ at PH symmetry after a quench of the local interaction. We set $T=0.1\Gamma$ and choose a fixed value of the gap $E_g=\Gamma$. We compare two kind of processes: one starting from $U_-=0$ to $U_+\neq0$ and the reverse one, which starts from $U_-\neq0$ and quenches to $U_+=0$. Quite interestingly we see that, provided the average work $\bar{W}$ is \emph{above} the threshold of the (semi)gap $E_g$, as for the process $U_-=0\rightarrow U_+=10\Gamma$ (black curve in top panel) for which $\vert\bar{W}\vert=2.5\Gamma$, a rather fast thermalization can occurr also in the gapped model. Notice indeed that the expected thermal value for $D$, which is set by the dashed line in figure~\ref{fig:fix_gap_tune_quench} (top panel) and which corresponds to the value of double occupation computed at equilibrium for $U_+=10\Gamma$, is approached on a rather short time scale. 
We compare these findings with the inverse quench process, starting from $U_-=10\Gamma$ and quenching to $U_+=0$, which in force of Eq. (\ref{eqn:work_local_U}) is characterized by a rather small average work $\bar{W}< E_g$. As we see in figure~\ref{fig:fix_gap_tune_quench} the dynamics looks much slower in this case and we cannot conclude, on the basis of our data, wheter the thermal plateau at $D_{therm}=1/4$ is actually approached or not at longer times. A similar comparative study is performed for quenches from $U_-=0$ to $U_+=2.5\Gamma$ and viceversa and the results are plotted in bottom panel of figure~\ref{fig:fix_gap_tune_quench}. 

\subsubsection{Pseudo-gapped Fermionic Reservoir}

\begin{figure}[t]
\begin{center}
\epsfig{figure=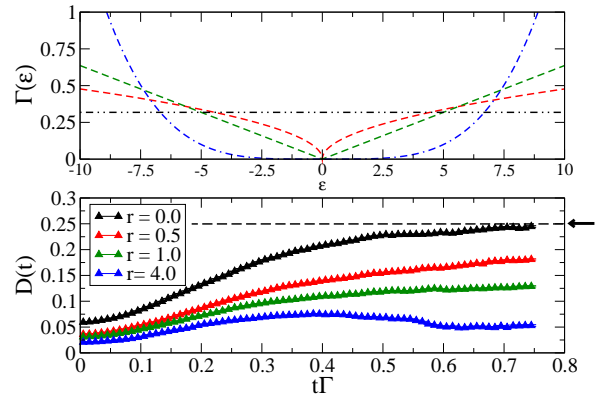,scale=0.3}
\caption{Non equilibrium dynamics for an Anderson Impurity coupled to a \emph{pseudo-gapped} fermionic reservoir. We consider the model at PH symmetry and $T=0.1\Gamma$. We fix the quench parameters, namely the initial and final value of the interaction, equal respectively to $U_-=10$ and $U_+=0$, and tune the pseudo-gap exponent from $r=0$ (gapless metallic state) to $r=4$. The depletion of low energy states in the DoS reflects into a much slower dynamics which eventually, for large enough $r$, seems to prevent the system from reaching the value of $D_{therm}=1/4$ which is fixed by PH symmetry and by the choice of $U_+=0$. However, due to finite time resolution we cannot conclude with present data wheter thermalization occurs or not on a longer time scale.}
\label{fig:fix_quench_tune_pwr} 
\end{center}
\end{figure}

We now consider the dynamics of an Anderson impurity coupled to a pseudogap reservoir. 
We consider as DoS a pure power-law function, namely
\be\label{eqn:band_pseudogap}
\rho_{pg}\left(\eps\right)=
\left\{
\begin{array}{lll}
\alpha\vert\eps\vert^{r} & &0<\vert\eps\vert< W \\
0& &\vert\eps\vert>W 
\end{array}
\right.\,,
\ee
where $\alpha=(r+1)/(2W^{r+1})$ ensures the proper normalization. This gives rise to a power-law hybridization function $\Gamma(\eps)$, that we define in complete analogy with the previous cases see Eq. (\ref{eqn:hybrid}). Again, we choose as unit of energy the hybridization width $\Gamma=\pi\,V^2/2W$.

The equilibrium phase diagram of the pseudo-gap Anderson Impurity model is extremely rich, featuring at particle-hole (PH) symmetry and for $0<r<1/2$, a quantum phase transition at a critical value of the hybridization $\Gamma_c$ between a strong coupling regime (for $\Gamma>\Gamma_c$) where Kondo screening occurs and a local moment one (for $\Gamma<\Gamma_c$) where the impurity becomes asymptotically free at low temperature. As opposite, at PH symmetry and for $r>1/2$ the only stable fixed point is the local moment one and no Kondo effect can be stabilized for an Anderson Impurity in a gapless reservoir\cite{Ingersent_PRB}.
In the following we will focus for simplicity on this latter case ($r>1/2$ at PH symmetry) so to avoid any complication related to the dynamics across criticality. We note that the topic of non equilibrium dynamics across quantum phase transitions is indeed an extremely intriguing problem\cite{Anders_Vojta_spinboson}, which may deserves furhter investigations in the future. However it goes well beyond the scope of the present paper. 

\begin{figure}[t]
\begin{center}
\epsfig{figure=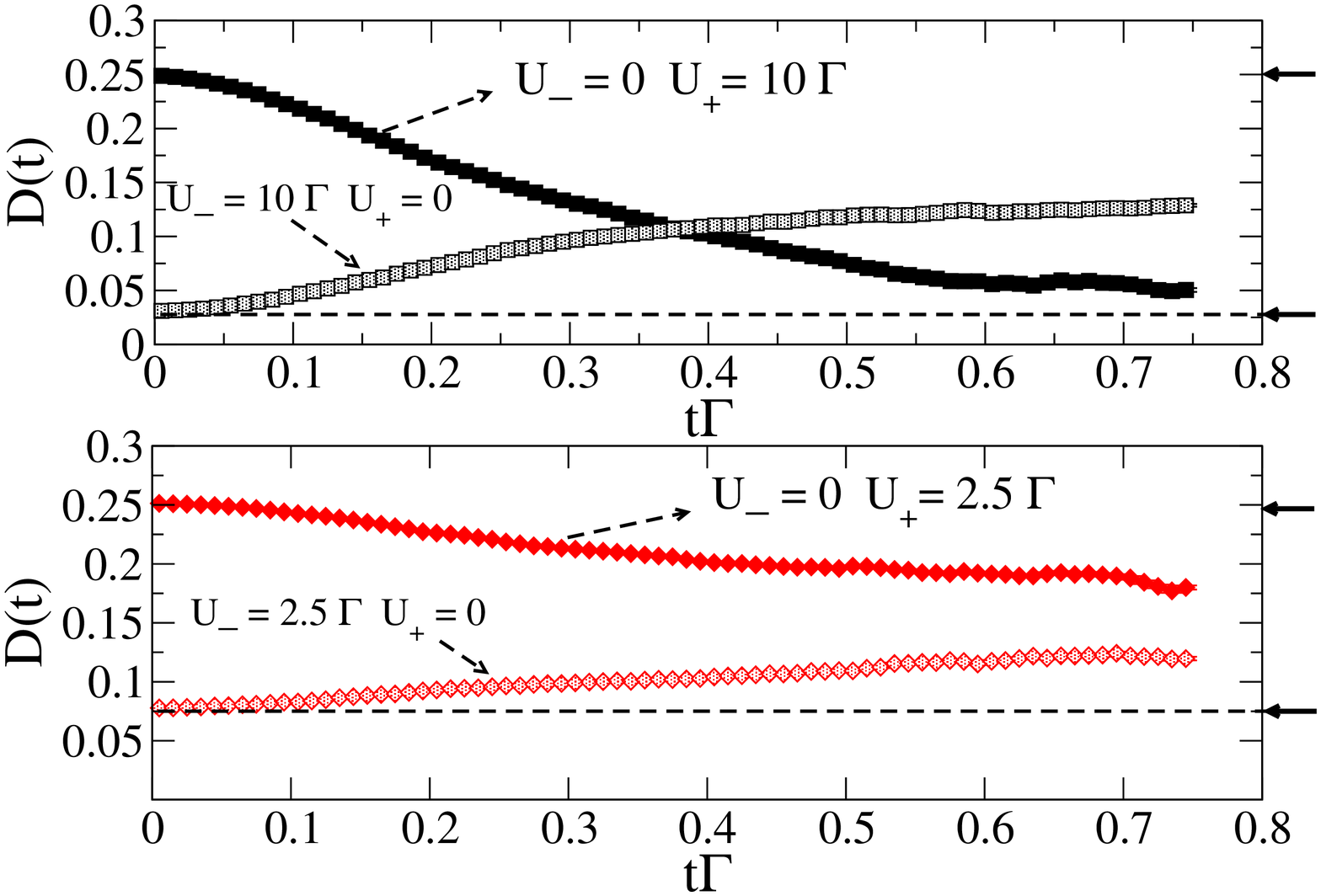,scale=0.3}
\caption{Non equilibrium dynamics for an Anderson Impurity coupled to a \emph{pseudo-gapped} fermionic reservoir. We consider the PH simmetric point, at $T=0.1\Gamma$ and for $r=1$. We study how the dynamics changes while tuning the average work $\bar{W}$ done during the quench (see text). We fix the initial interaction $U_-=0$ and perform a quench to $U_+=10\Gamma$ (top panel). We compare the results with the reverse process, from $U_-\neq0$ to $U_+=0$. Despite the power-law DoS, the dynamics in the case of a large quantum quench (large average work) turns to be quite fast. On the contrary, in the low work regime (bottom panel) the dynamics is much slower and we cannot see wheter thermalization take place or not at longer time scales. }
\label{fig:fix_pwr_tune_quench} 
\end{center}
\end{figure}
As we did for the gapped case, we consider as a starting point a quantum quench of the local Coulomb interaction between $U_-=10\Gamma$ and $U_+=0$, at $T=0.1\Gamma$ and for $\eps_{d+}=\eps_{d-}=0$. 
In figure~\ref{fig:fix_quench_tune_pwr} we plot the dynamics of double occupancy as a function of time, while tuning the exponent $r$ from the metallic case $r=0$ to the almost gapped one $r=4$. As we can see the results we found look very similar to what already discussed in the case of a gapped reservoir. While for $r=0$, namely for a finite hybridization at the Fermi level, the dynamics is pretty fast in approaching the thermal plateau (we are at PH simmetry and $U_+=0$, therefore again $D_{therm}=1/4$), for $r\neq0$ the dynamics slows down and for $r=4$ seems actually to get stucked into a non thermal steady state. However from these data it is difficult to conclude wheter this is really the case or rather that thermalization emerges on a very long time scale.

We conclude this section by discussing how the dynamics in the pseudo-gapped case changes as a function of the work done during the quantum quench. To this extent we plot in figure~\ref{fig:fix_pwr_tune_quench} the double occupancy as a function of time, $D(t)$, for $T=0.1\Gamma$ and at PH symmetry. As we have previously done, we fix the initial value of the interaction to $U_-=0$ while tuning the final value $U_+$ (see top of panel) so to change the average work $\bar{W}$. At the same time, we study the dynamics for the reverse process where we fix the final value of the repulsion to $U_+=0$, while changing the initial condition $U_-$. As we have already found in the gapped case, the dynamics turns out to be very sensitive to the average work done, namely to the amount of energy pushed into the system. In particular we can see from figure~\ref{fig:fix_pwr_tune_quench} that quenches with sufficiently large work $\bar{W}$ can result into a rather fast dynamics and thermalization at long times. This seems to be the case, for example, of quenches  from $U_-=0$ to $U_+=10\Gamma$ (black points, top panel) where the thermal value of double occupation with $U_+=10\Gamma$ is set by the arrow at the bottom of the panel. In other cases, where the work done is not that large, the dynamics turns to be slow and we cannot conclude about the long time behaviour.

\subsubsection{Discussion}

To summarize, in this section we have discussed the non equilibrium quench dynamics of the Anderson Impurity model in a gapped or pseudo-gapped fermionic reservoir after a quantum quench of the local Coulomb interaction. For the sake of simplicity we have considered only the particle-hole simmetric case and we have chosen the parameters in such a way to be always in the local moment regime in equilibrium for both gapped and pseudo-gapped cases, so to avoid further complications due to local quantum criticality which may result into very low-energy/long-time scales controlling the dynamics.

An important issue we have tried to discuss concerns the onset of thermalization at long times. While this is expected to occurr for quantum quenches in a conventional metallic reservoir, one may wonder wheter or not the lack of available states close to the Fermi energy could result into a lack of thermalization at long times.
We have shown that the real-time dynamics is strongly affected, even on short-to-intermediate time scales, by the modified spectral function of the bath. In particular, the opening of a gap or pseudo-gap at the Fermi level results into a slower transient dynamics. While it is tempting to explain this fact by invoking the equilibrium properies of the model and the mentioned flow to the local moment fixed point, one have to take into account also the intrinsic out-of-equilibrium nature of the quantum quench process. In this perspective we have identified the work $\bar{W}$ done during the quench as a relevant physical quantity to describe the non equilibrium dynamics after the quench. In particular, for the gapped and pseudogapped models, we have shown that a rather fast thermalization can occur provided the work done $\bar{W}$ is sufficiently large (for example with respect to the (semi)gap $E_g$). As opposite for small quantum quenches characterized by a small amount of work done $\bar{W}\ll E_g$, the dynamics turns to be much slower and we cannot conclude, with present data, wheter thermalization occurs or not on a longer time scale, thus leaving the question open to further investigations.

\section{Conclusions}

In this work we have extended the recently developed real-time diagMC method, in its hybridization expansion formulation, to full 
Kadanoff-Baym-Keldysh contour. The resulting algorithm represents a completely general and numerically exact approach 
to real-time dynamics in quantum impurity models, which interpolates between the standard equilibrium diagMC\cite{ctqmc_Werner} 
defined on the Matsubara imaginary-time axis and its recent non equilibrium extensions\cite{Keldysh_short,Rabani,Werner_Keldysh_09} 
that works on the Keldysh contour and requires a special choice of the initial condition for the dynamics.
Merging together these two approaches, we are able to deal with the most generic setup, namely a strongly correlated quantum 
impurity model initially in thermal equilibrium, which is driven out of equilibrium by some external time dependent perturbation. 
As a consequence, several kind of initial preparations as well as driving protocols can be considered with our approach that allow 
studying a wide class of non equilibrium problems. 

More interestingly, we notice that no constraints are required on the nature of the fermionic \emph{bath}, which enters in our 
approach as an input, encoded in the contour-ordered hybridization function $i\Delta_C\left(t,t'\right)$. This allows us to deal 
with the intriguing problem of studying the real-time dynamics of the quantum impurity coupled to a non equilibrium bath and 
opens the way to applying our method to solve Non Equilibrium Dynamical Mean Field Theory. In this perspective the effective action formulation of the algorithm we have presented at the end of section \ref{sect:hybrid_exp} represents the most natural one. The natural extension of this research is to study relaxation dynamics in correlated macroscopic quantum systems using Non Equilibrium Dynamical Mean Field Theory.

As a first application of our algorithm we have studied the real-time in an Anderson Impurity Model after a local quantum quench. In the case of a metallic reservoir we have discussed time scales controlling charge and spin relaxation. While the former is a rather fast process mainly controlled by hybridization $\Gamma$, the latter turns to be a much slower process associated with the lowest energy scale in the problem, namely the Kondo temperature $T_K$. As we have shown in section \ref{sec:charge_spin} , the charge time scale can be reached within the present approach, while the decay of a polarized spin cannot, due to sign problem which makes calculations at very long times increasingly difficult. 

Finally we have addressed the non equilibrium dynamics of an Anderson Impurity coupled to a gapped or a pseudo-gapped reservoir. Even though we restrict our attention to the PH symmetry and to power-law exponents $r>1/2$, for which the equilibrium phase diagram in both gapped and pseudo-gapped cases only features a local moment fixed point, the real-time dynamics for charge degrees of freedom turns out to be rather intriguing. In particular we distinguish two regimes depending on wheter the amount of work done during the quench, $\bar{W}$, is large or small with respect to typical energy scale in the DoS. In the former case we observe a rather fast dynamics which may give rise to thermalization, while in the latter case a much slower dynamics which prevent us from drawing definite conclusions on the long-time behaviour. The investigation of real-time dynamics in this class of quantum impurity models represents, in this perspective, a very intriguing and challenging open problem.

\acknowledgments
I am grateful to Michele Fabrizio and Massimo Capone for many stimulating discussions, continuous support and for a carefull reading of this manuscript, and to Erio Tosatti for discussions and support during this project. Allocation of computational resources from Caspur under the Standard HPC grant std09-419 is also acknowledged.
\appendix

\section{Contour-ordered Hybridization Function}\label{app:contour_ord_hyb}

In this appendix we discuss with some more details the contour-ordered Hybridization Function $i\Delta_C\left(t,t'\right)$ we have introduced in the text,  which is the basic object entering the hybridization expansion on the Kadanoff-Baym-Keldysh contour $\m{C}$. This function encodes the effect of the bath on the impurity degrees of freedom, as it clearly appears in the effective action formulation of the theory. 
As we have shown in section \ref{sect:hybrid_exp}, in the case of a quantum impurity coupled to an equilibrium fermionic bath the hybridization function $i\Delta_C\left(t_1,t_2\right)$ can be written as
\be\label{eqn:delta_contour_app}
i\Delta_C\left(t_1,t_2\right)\equiv 
\sum_{\mathbf{k}}\,V^2_{\mathbf{k}}\,\langle\,T_C\,f_{\mathbf{k}}\left(t_1\right)
f^{\dagger}_{\mathbf{k}}\left(t_2\right)\rangle_{bath}\,,
\ee
while in general, i.e. for out of equilibrium fermionic baths, a parametrization of this function in terms of time independent Anderson impurity Hamiltonian is not possible. From the previous expression we see that 
$i\Delta_C\left(t_1,t_2\right)$ is given by the contour-ordered bath Green's function evaluated at the impurity site. The meaning of this function is the following. We consider a bath of free fermionic excitations in equilibrium at temperature $T$, whose hamiltonian generally reads
\be\label{eqn:H_bath}
\m{H}_{bath}=
\sum_{\mathbf{k}\,a}\,\varepsilon_{\mathbf{k}}\,\fdag_{\mathbf{k}\,a}\,f_{\mathbf{k}\,a}
\ee
We take as initial density matrix $\rho_{0}$ the statistical one, 
\be\label{eqn:rho_init_bath}
\rho_{in}=\frac{e^{-\beta\,\m{H}_{bath}}}{Z}\,,
\ee
and define the contour-ordered bath Green's function as
\be\label{eqn:contour_green_bath}
g_{\mb{k}\,a}\left(t,t'\right)=-i\langle\,T_{\m{C}}\,\left(f_{\mathbf{k}\,a}\left(t\right)\fdag_{\mathbf{k}\,a}\left(t'\right)
\right)\,\rangle\,,\qquad t,t'\in \m{C}
\ee
where both time arguments $t$ and $t'$ live on the three branch contour $\m{C}$, while the average is taken over the initial density matrix. The contour time ordering $T_{\m{C}}$ acts as described in the main text, namely ordering operators according to their time argument on the contour $\m{C}$. Concerning the contour-time evolution of creation and annihilation operators it is defined as usual
\be\label{eqn:heisen_pict}
f_{\mathbf{k}\,a}\left(t\right)\,=\,e^{i\m{H}_{bath}t}\,f_{\mathbf{k}\,a}\,e^{-i\m{H}_{bath}t}\,.
\ee

We now discuss the possible time orderings arising from the choice of $t$ and $t'$ along the contour $\m{C}$, which naturally lead to a $3\times 3$ matrix structure for the contour-ordered hybridizaton function.
\be\label{eqn:delta_contour_3x3_app}
i\Delta_C\left(t,t'\right)\rightarrow\;i\Delta_{ab}\left(t,t'\right)
\,\qquad a,b=1,2,3
\ee
\subsection{Matsubara Sector}

When both arguments live on the imaginary time axis, namely $t=-i\tau$ and $t'=-i\tau'$, we recover the standard Matsubara Green's function, a part from the $i$-factor
\bea
&g^{33}_{\mb{k}\,a}\left(\tau,\tau'\right)=-i\langle
T_{\tau}\,\left(f_{\mathbf{k}\,a}\left(\tau\right)\fdag_{\mathbf{k}\,a}\left(\tau'\right)
\right)\,\rangle\,.&
\eea
We note this function is antiperiodic and time-traslational invariant, therefore we can set $\tau'=0$ and compute it in the interval $\tau\in\left[0,\beta\right]$. We obtain for the hybridization function the standard result used also in equilibrium diagMC
\be
\Delta_{33}\left(\tau\right) = -i\,\int\,\frac{d\eps}{\pi}\,\Gamma\left(\eps\right)n\left(\eps\right)\,e^{-\eps\tau}\,,
\ee
where $n\left(\eps\right)$ is the Fermi distribution function and we had explicitly introduced the energy-dependent the hybridization $\Gamma\left(\eps\right)$ 
\be
\Gamma(\eps)=\pi\,\sum_{k}\,\vert V_{k}\vert^2\,\delta(\eps-\eps_k)\,.
\ee
\subsection{Keldysh Sector}

When $t$ and $t'$ are both on real-time branches we are in the Keldysh subspace. The Green's function acquires a $2\times 2$ matrix form, depending on the branch position ($a,b=1,2$) of the two time arguments, and consequently also the hybridization function in Eq. (\ref{eqn:delta_contour_app}) can be written as a matrix
\bea
\Delta_{ab}\left(t,t'\right)=
\left(
\begin{array}{cc}
\Delta^{11}\left(t,t'\right)  & \Delta^{12}\left(t,t'\right)\\
\Delta^{21}\left(t,t'\right) & \Delta^{22}\left(t,t'\right) \\
\end{array}
\right)\,.
\eea
In particular, when $t$ is greater/lesser than $t'$ on the contour we get the $\Delta^{21}$/$\Delta^{12}$ component, which reads respectively
\be\label{eqn:greater_gf_app}
\Delta^{21}\left(t,t'\right)=
\int\,\frac{d\eps}{\pi}\,\Gamma\left(\eps\right)\,\left(1-n_F\left(\eps\right)\right)\,
e^{-i\,\eps\,\left(t-t'\right)}
\ee
and
\be\label{eqn:lesser_gf_app}
\Delta^{12}\left(t,t'\right)=
-\int\,\frac{d\eps}{\pi}\,\Gamma\left(\eps\right)\,n_F\left(\eps\right)\,
e^{-i\,\eps\,\left(t-t'\right)}\,.
\ee
On the contrary when the two times are on the same branch we obtain the time-ordered  $\Delta^{11}$ or anti-time ordered $\Delta^{12}$ hybridization function, which reduce to the off-diagonal ones depending on the time interval, namely
\be\label{eqn:diagonal_11_gf_app}
\Delta^{11}\left(t,t'\right)=\left\{
\begin{array}{ll}
 t > t' & \Delta^{21}\left(t,t'\right)\\
 t < t' & \Delta^{12}\left(t,t'\right)
\end{array}
\right.
\ee
and
\be\label{eqn:diagonal_22_gf_app}
\Delta^{22}\left(t,t'\right)=\left\{
\begin{array}{ll}
 t > t' & \Delta^{12}\left(t,t'\right)\\
 t < t' & \Delta^{21}\left(t,t'\right)
\end{array}
\right.
\ee
We note that, by construction, since we are considering a time independent quantum impurity model, all the four Keldysh components only depend on time differences $t-t'$. This property is peculiar of a bath which is in thermal equilibrium and do not hold in general for other kinds of non equilibrium driving protocols or in the case we are solving Non-Equilibrium DMFT.

\subsection{Mixed Sector}

Finally we have to consider the case in which the two time arguments live in different sectors. These mixed hybridization functions usually take into account the short-time memory effects, namely the transient correlations due to the chosen initial density matrix. We can distinguish two cases, depending on wheter $t>_C t'$ or viceversa. In the former case, namely when $t = -i\tau$ is on the Matsubara branch while $t'$ lies on the Keldysh branches we have                                                                                                                                                                                                                                                                   
\be\label{eqn:mix_gf_31_app}
\Delta_{31}\left(\tau,t'\right)=
-\,\int\,\frac{d\eps}{\pi}\,\Gamma\left(\eps\right)\,\left(1-n_F\left(\eps\right)\right)\,
e^{-\eps\,\tau}\,e^{-i\,\eps\,t'}\,
\ee
as well as
\be\label{eqn:mix_gf_23_app}
\Delta_{32}\left(\tau,t'\right)=
-\,\int\,\frac{d\eps}{\pi}\,\Gamma\left(\eps\right)\,\left(1-n_F\left(\eps\right)\right)\,
e^{-\eps\,\tau}\,e^{-i\,\eps\,t'}\,.
\ee
We note that in this case, as usual for off-diagonal terms, the contour-time ordering is fixed independently from the values of the time arguments since the Keldysh branches are always lesser on the contour $\m{C}$ than the imaginary branch. Moreover, we note there is no difference in the value of the hybridization function if the real-time is placed on the upper or lower branch, namely
\be
\Delta^{31}\left(\tau,t'\right)=\Delta_{32}\left(\tau,t'\right)\,.
\ee.
As opposite, when $t$ lies on the Keldysh contour while $t' = -i\tau'$ is on the Matsubara branch we obtain
the other two mixed components
\be\label{eqn:mix_gf_13_app}
\Delta^{13}\left(t,\tau'\right)=
-\,\int\,\frac{d\eps}{\pi}\,\Gamma\left(\eps\right)\,n_F\left(\eps\right)\,
e^{\eps\,\tau'}\,e^{-i\,\eps\,t}\,=
\Delta^{23}\left(t,\tau'\right)\,.
\ee


\begin{thebibliography}{57}
\expandafter\ifx\csname natexlab\endcsname\relax\def\natexlab#1{#1}\fi
\expandafter\ifx\csname bibnamefont\endcsname\relax
  \def\bibnamefont#1{#1}\fi
\expandafter\ifx\csname bibfnamefont\endcsname\relax
  \def\bibfnamefont#1{#1}\fi
\expandafter\ifx\csname citenamefont\endcsname\relax
  \def\citenamefont#1{#1}\fi
\expandafter\ifx\csname url\endcsname\relax
  \def\url#1{\texttt{#1}}\fi
\expandafter\ifx\csname urlprefix\endcsname\relax\def\urlprefix{URL }\fi
\providecommand{\bibinfo}[2]{#2}
\providecommand{\eprint}[2][]{\url{#2}}

\bibitem[{\citenamefont{Cavalieri et~al.}(2007)\citenamefont{Cavalieri,
  M\"{u}ller, Uphues, Yakovlev, Baltuska, Horvath, Schmidt, Bl\"{u}mel,
  Holzwarth, Hendel et~al.}}]{Cavalieri_nature06}
\bibinfo{author}{\bibfnamefont{A.~L.} \bibnamefont{Cavalieri}},
  \bibinfo{author}{\bibfnamefont{N.}~\bibnamefont{M\"{u}ller}},
  \bibinfo{author}{\bibfnamefont{T.}~\bibnamefont{Uphues}},
  \bibinfo{author}{\bibfnamefont{V.~S.} \bibnamefont{Yakovlev}},
  \bibinfo{author}{\bibfnamefont{A.}~\bibnamefont{Baltuska}},
  \bibinfo{author}{\bibfnamefont{B.}~\bibnamefont{Horvath}},
  \bibinfo{author}{\bibfnamefont{B.}~\bibnamefont{Schmidt}},
  \bibinfo{author}{\bibfnamefont{L.}~\bibnamefont{Bl\"{u}mel}},
  \bibinfo{author}{\bibfnamefont{R.}~\bibnamefont{Holzwarth}},
  \bibinfo{author}{\bibfnamefont{S.}~\bibnamefont{Hendel}},
  \bibnamefont{et~al.}, \bibinfo{journal}{Nature}
  \textbf{\bibinfo{volume}{449}}, \bibinfo{eid}{1029} (\bibinfo{year}{2007}).

\bibitem[{\citenamefont{Zewail}(2000)}]{Zewail_jchem_00}
\bibinfo{author}{\bibfnamefont{A.~H.} \bibnamefont{Zewail}},
  \bibinfo{journal}{The Journal of Physical Chemistry A}
  \textbf{\bibinfo{volume}{104}}, \bibinfo{eid}{5660} (\bibinfo{year}{2000}).

\bibitem[{\citenamefont{Perfetti et~al.}(2006)\citenamefont{Perfetti, Loukakos,
  Lisowski, Bovensiepen, Berger, Biermann, Cornaglia, Georges, and
  Wolf}}]{Perfetti_Georges_PRL06}
\bibinfo{author}{\bibfnamefont{L.}~\bibnamefont{Perfetti}},
  \bibinfo{author}{\bibfnamefont{P.~A.} \bibnamefont{Loukakos}},
  \bibinfo{author}{\bibfnamefont{M.}~\bibnamefont{Lisowski}},
  \bibinfo{author}{\bibfnamefont{U.}~\bibnamefont{Bovensiepen}},
  \bibinfo{author}{\bibfnamefont{H.}~\bibnamefont{Berger}},
  \bibinfo{author}{\bibfnamefont{S.}~\bibnamefont{Biermann}},
  \bibinfo{author}{\bibfnamefont{P.~S.} \bibnamefont{Cornaglia}},
  \bibinfo{author}{\bibfnamefont{A.}~\bibnamefont{Georges}}, \bibnamefont{and}
  \bibinfo{author}{\bibfnamefont{M.}~\bibnamefont{Wolf}},
  \bibinfo{journal}{Phy. Rev. Lett.} \textbf{\bibinfo{volume}{97}},
  \bibinfo{eid}{067402} (\bibinfo{year}{2006}).

\bibitem[{\citenamefont{Giannetti et~al.}(2009)\citenamefont{Giannetti,
  Coslovich, Cilento, Ferrini, Eisaki, Kaneko, Greven, and
  Parmigiani}}]{Giannetti_PRB09_BISCCO}
\bibinfo{author}{\bibfnamefont{C.}~\bibnamefont{Giannetti}},
  \bibinfo{author}{\bibfnamefont{G.}~\bibnamefont{Coslovich}},
  \bibinfo{author}{\bibfnamefont{F.}~\bibnamefont{Cilento}},
  \bibinfo{author}{\bibfnamefont{G.}~\bibnamefont{Ferrini}},
  \bibinfo{author}{\bibfnamefont{H.}~\bibnamefont{Eisaki}},
  \bibinfo{author}{\bibfnamefont{N.}~\bibnamefont{Kaneko}},
  \bibinfo{author}{\bibfnamefont{M.}~\bibnamefont{Greven}}, \bibnamefont{and}
  \bibinfo{author}{\bibfnamefont{F.}~\bibnamefont{Parmigiani}},
  \bibinfo{journal}{Phys. Rev. B} \textbf{\bibinfo{volume}{79}},
  \bibinfo{eid}{224502} (\bibinfo{year}{2009}).

\bibitem[{\citenamefont{Greiner et~al.}(2002)\citenamefont{Greiner, Mandel,
  Esslinger, Hansch, and Bloch}}]{Greiner_nature_02}
\bibinfo{author}{\bibfnamefont{M.}~\bibnamefont{Greiner}},
  \bibinfo{author}{\bibfnamefont{O.}~\bibnamefont{Mandel}},
  \bibinfo{author}{\bibfnamefont{T.}~\bibnamefont{Esslinger}},
  \bibinfo{author}{\bibfnamefont{T.}~\bibnamefont{Hansch}}, \bibnamefont{and}
  \bibinfo{author}{\bibfnamefont{I.}~\bibnamefont{Bloch}},
  \bibinfo{journal}{Nature} \textbf{\bibinfo{volume}{419}}, \bibinfo{eid}{51}
  (\bibinfo{year}{2002}).

\bibitem[{\citenamefont{Strohmaier et~al.}(2009)\citenamefont{Strohmaier,
  Greif, Jordens, Tarruell, Moritz, Esslinger, Sensarma, Pekker, Altman, and
  Demler}}]{Doublon_decay}
\bibinfo{author}{\bibfnamefont{N.}~\bibnamefont{Strohmaier}},
  \bibinfo{author}{\bibfnamefont{D.}~\bibnamefont{Greif}},
  \bibinfo{author}{\bibfnamefont{R.}~\bibnamefont{Jordens}},
  \bibinfo{author}{\bibfnamefont{L.}~\bibnamefont{Tarruell}},
  \bibinfo{author}{\bibfnamefont{H.}~\bibnamefont{Moritz}},
  \bibinfo{author}{\bibfnamefont{T.}~\bibnamefont{Esslinger}},
  \bibinfo{author}{\bibfnamefont{R.}~\bibnamefont{Sensarma}},
  \bibinfo{author}{\bibfnamefont{D.}~\bibnamefont{Pekker}},
  \bibinfo{author}{\bibfnamefont{E.}~\bibnamefont{Altman}}, \bibnamefont{and}
  \bibinfo{author}{\bibfnamefont{E.}~\bibnamefont{Demler}},
  \bibinfo{journal}{arXiv:0905.2963}  (\bibinfo{year}{2009}).

\bibitem[{\citenamefont{Trotzky et~al.}(2006)\citenamefont{Trotzky, Cheinet,
  F\"{o}lling, Feld, Schnorrberger, Rey, Polkovnikov, Demler, Lukin, and
  Bloch}}]{Superexchange_science}
\bibinfo{author}{\bibfnamefont{S.}~\bibnamefont{Trotzky}},
  \bibinfo{author}{\bibfnamefont{P.}~\bibnamefont{Cheinet}},
  \bibinfo{author}{\bibfnamefont{S.}~\bibnamefont{F\"{o}lling}},
  \bibinfo{author}{\bibfnamefont{M.}~\bibnamefont{Feld}},
  \bibinfo{author}{\bibfnamefont{U.}~\bibnamefont{Schnorrberger}},
  \bibinfo{author}{\bibfnamefont{A.~M.} \bibnamefont{Rey}},
  \bibinfo{author}{\bibfnamefont{A.}~\bibnamefont{Polkovnikov}},
  \bibinfo{author}{\bibfnamefont{E.~A.} \bibnamefont{Demler}},
  \bibinfo{author}{\bibfnamefont{M.~D.} \bibnamefont{Lukin}}, \bibnamefont{and}
  \bibinfo{author}{\bibfnamefont{I.}~\bibnamefont{Bloch}},
  \bibinfo{journal}{Science} \textbf{\bibinfo{volume}{319}}
  (\bibinfo{year}{2006}).

\bibitem[{\citenamefont{Vandersypen et~al.}(2004)\citenamefont{Vandersypen,
  Elzerman, Schouten, van Beveren, Hanson, and Kouwenhoven}}]{Elzerman_APL_04}
\bibinfo{author}{\bibfnamefont{L.~M.~K.} \bibnamefont{Vandersypen}},
  \bibinfo{author}{\bibfnamefont{J.~M.} \bibnamefont{Elzerman}},
  \bibinfo{author}{\bibfnamefont{R.~N.} \bibnamefont{Schouten}},
  \bibinfo{author}{\bibfnamefont{L.~H.~W.} \bibnamefont{van Beveren}},
  \bibinfo{author}{\bibfnamefont{R.}~\bibnamefont{Hanson}}, \bibnamefont{and}
  \bibinfo{author}{\bibfnamefont{L.~P.} \bibnamefont{Kouwenhoven}},
  \bibinfo{journal}{Applied Physics Letters} \textbf{\bibinfo{volume}{85}}
  (\bibinfo{year}{2004}).

\bibitem[{\citenamefont{Schleser et~al.}(2004)\citenamefont{Schleser, Ruh, Ihn,
  Ensslin, Driscoll, and Gossard}}]{Ensslin_APL_04}
\bibinfo{author}{\bibfnamefont{R.}~\bibnamefont{Schleser}},
  \bibinfo{author}{\bibfnamefont{E.}~\bibnamefont{Ruh}},
  \bibinfo{author}{\bibfnamefont{T.}~\bibnamefont{Ihn}},
  \bibinfo{author}{\bibfnamefont{K.}~\bibnamefont{Ensslin}},
  \bibinfo{author}{\bibfnamefont{D.~C.} \bibnamefont{Driscoll}},
  \bibnamefont{and} \bibinfo{author}{\bibfnamefont{A.~C.}
  \bibnamefont{Gossard}}, \bibinfo{journal}{Applied Physics Letters}
  \textbf{\bibinfo{volume}{85}} (\bibinfo{year}{2004}).

\bibitem[{\citenamefont{Gustavsson et~al.}(2009)\citenamefont{Gustavsson,
  Leturcq, Studer, Shorubalko, Ihn, Ensslin, Driscoll, and
  Gossard}}]{Gustavsson_2009}
\bibinfo{author}{\bibfnamefont{S.}~\bibnamefont{Gustavsson}},
  \bibinfo{author}{\bibfnamefont{R.}~\bibnamefont{Leturcq}},
  \bibinfo{author}{\bibfnamefont{M.}~\bibnamefont{Studer}},
  \bibinfo{author}{\bibfnamefont{I.}~\bibnamefont{Shorubalko}},
  \bibinfo{author}{\bibfnamefont{T.}~\bibnamefont{Ihn}},
  \bibinfo{author}{\bibfnamefont{K.}~\bibnamefont{Ensslin}},
  \bibinfo{author}{\bibfnamefont{D.}~\bibnamefont{Driscoll}}, \bibnamefont{and}
  \bibinfo{author}{\bibfnamefont{A.}~\bibnamefont{Gossard}},
  \bibinfo{journal}{Surface Science Reports} \textbf{\bibinfo{volume}{64}},
  \bibinfo{pages}{191 } (\bibinfo{year}{2009}).

\bibitem[{\citenamefont{Wilson}(1975)}]{Wilson_RMP}
\bibinfo{author}{\bibfnamefont{K.~G.} \bibnamefont{Wilson}},
  \bibinfo{journal}{Rev. Mod. Phys.} \textbf{\bibinfo{volume}{47}},
  \bibinfo{pages}{773} (\bibinfo{year}{1975}).

\bibitem[{\citenamefont{Bulla et~al.}(2008)\citenamefont{Bulla, Costi, and
  Pruschke}}]{Pruschke_RMP}
\bibinfo{author}{\bibfnamefont{R.}~\bibnamefont{Bulla}},
  \bibinfo{author}{\bibfnamefont{T.~A.} \bibnamefont{Costi}}, \bibnamefont{and}
  \bibinfo{author}{\bibfnamefont{T.}~\bibnamefont{Pruschke}},
  \bibinfo{journal}{Rev. Mod. Phys.} \textbf{\bibinfo{volume}{80}},
  \bibinfo{eid}{395} (\bibinfo{year}{2008}).

\bibitem[{\citenamefont{Anders and Schiller}(2005)}]{tnrg_Anders}
\bibinfo{author}{\bibfnamefont{F.~B.} \bibnamefont{Anders}} \bibnamefont{and}
  \bibinfo{author}{\bibfnamefont{A.}~\bibnamefont{Schiller}},
  \bibinfo{journal}{Phys. Rev. Lett.} \textbf{\bibinfo{volume}{95}},
  \bibinfo{eid}{196801} (\bibinfo{year}{2005}).

\bibitem[{\citenamefont{Heidrich-Meisner
  et~al.}(2009)\citenamefont{Heidrich-Meisner, Feiguin, and
  Dagotto}}]{DMRG_Dagotto_09}
\bibinfo{author}{\bibfnamefont{F.}~\bibnamefont{Heidrich-Meisner}},
  \bibinfo{author}{\bibfnamefont{A.~E.} \bibnamefont{Feiguin}},
  \bibnamefont{and} \bibinfo{author}{\bibfnamefont{E.}~\bibnamefont{Dagotto}},
  \bibinfo{journal}{Phys. Rev. B} \textbf{\bibinfo{volume}{79}},
  \bibinfo{eid}{235336} (pages~\bibinfo{numpages}{6}) (\bibinfo{year}{2009}).

\bibitem[{\citenamefont{Guo et~al.}(2009)\citenamefont{Guo, Weichselbaum,
  Kehrein, Xiang, and von Delft}}]{Weichselbaum_PRB_09}
\bibinfo{author}{\bibfnamefont{C.}~\bibnamefont{Guo}},
  \bibinfo{author}{\bibfnamefont{A.}~\bibnamefont{Weichselbaum}},
  \bibinfo{author}{\bibfnamefont{S.}~\bibnamefont{Kehrein}},
  \bibinfo{author}{\bibfnamefont{T.}~\bibnamefont{Xiang}}, \bibnamefont{and}
  \bibinfo{author}{\bibfnamefont{J.}~\bibnamefont{von Delft}},
  \bibinfo{journal}{Phys. Rev. B} \textbf{\bibinfo{volume}{79}},
  \bibinfo{eid}{115137} (\bibinfo{year}{2009}).

\bibitem[{\citenamefont{Weiss et~al.}(2008)\citenamefont{Weiss, Eckel,
  Thorwart, and Egger}}]{iter_Egger}
\bibinfo{author}{\bibfnamefont{S.}~\bibnamefont{Weiss}},
  \bibinfo{author}{\bibfnamefont{J.}~\bibnamefont{Eckel}},
  \bibinfo{author}{\bibfnamefont{M.}~\bibnamefont{Thorwart}}, \bibnamefont{and}
  \bibinfo{author}{\bibfnamefont{R.}~\bibnamefont{Egger}},
  \bibinfo{journal}{Phys. Rev. B} \textbf{\bibinfo{volume}{77}},
  \bibinfo{eid}{195316} (\bibinfo{year}{2008}).

\bibitem[{\citenamefont{Hackl and Kehrein}(2008)}]{Kehrein_PRB_08}
\bibinfo{author}{\bibfnamefont{A.}~\bibnamefont{Hackl}} \bibnamefont{and}
  \bibinfo{author}{\bibfnamefont{S.}~\bibnamefont{Kehrein}},
  \bibinfo{journal}{Phys. Rev. B} \textbf{\bibinfo{volume}{78}},
  \bibinfo{eid}{092303} (\bibinfo{year}{2008}).

\bibitem[{\citenamefont{Georges et~al.}(1996)\citenamefont{Georges, Kotliar,
  Krauth, and Rozenberg}}]{Review_DMFT_96}
\bibinfo{author}{\bibfnamefont{A.}~\bibnamefont{Georges}},
  \bibinfo{author}{\bibfnamefont{G.}~\bibnamefont{Kotliar}},
  \bibinfo{author}{\bibfnamefont{W.}~\bibnamefont{Krauth}}, \bibnamefont{and}
  \bibinfo{author}{\bibfnamefont{M.~J.} \bibnamefont{Rozenberg}},
  \bibinfo{journal}{Rev. Mod. Phys.} \textbf{\bibinfo{volume}{68}},
  \bibinfo{pages}{13} (\bibinfo{year}{1996}).

\bibitem[{\citenamefont{Freericks et~al.}(2006)\citenamefont{Freericks,
  Turkowski, and Zlati\'{c}}}]{Non_Eq_DMFT}
\bibinfo{author}{\bibfnamefont{J.~K.} \bibnamefont{Freericks}},
  \bibinfo{author}{\bibfnamefont{V.~M.} \bibnamefont{Turkowski}},
  \bibnamefont{and}
  \bibinfo{author}{\bibfnamefont{V.}~\bibnamefont{Zlati\'{c}}},
  \bibinfo{journal}{Phys. Rev. Lett.} \textbf{\bibinfo{volume}{97}},
  \bibinfo{eid}{266408} (\bibinfo{year}{2006}).

\bibitem[{\citenamefont{Eckstein et~al.}(2009)\citenamefont{Eckstein, Kollar,
  and Werner}}]{Werner_DMFT}
\bibinfo{author}{\bibfnamefont{M.}~\bibnamefont{Eckstein}},
  \bibinfo{author}{\bibfnamefont{M.}~\bibnamefont{Kollar}}, \bibnamefont{and}
  \bibinfo{author}{\bibfnamefont{P.}~\bibnamefont{Werner}},
  \bibinfo{journal}{Phys. Rev. Lett.} \textbf{\bibinfo{volume}{103}},
  \bibinfo{eid}{056403} (\bibinfo{year}{2009}).

\bibitem[{\citenamefont{Rubtsov et~al.}(2005)\citenamefont{Rubtsov, Savkin, and
  Lichtenstein}}]{ctqmc_Rubtsov}
\bibinfo{author}{\bibfnamefont{A.~N.} \bibnamefont{Rubtsov}},
  \bibinfo{author}{\bibfnamefont{V.~V.} \bibnamefont{Savkin}},
  \bibnamefont{and} \bibinfo{author}{\bibfnamefont{A.~I.}
  \bibnamefont{Lichtenstein}}, \bibinfo{journal}{Phys. Rev. B}
  \textbf{\bibinfo{volume}{72}}, \bibinfo{eid}{035122} (\bibinfo{year}{2005}).

\bibitem[{\citenamefont{Werner et~al.}(2006)\citenamefont{Werner, Comanac, de'
  Medici, Troyer, and Millis}}]{ctqmc_Werner}
\bibinfo{author}{\bibfnamefont{P.}~\bibnamefont{Werner}},
  \bibinfo{author}{\bibfnamefont{A.}~\bibnamefont{Comanac}},
  \bibinfo{author}{\bibfnamefont{L.}~\bibnamefont{de' Medici}},
  \bibinfo{author}{\bibfnamefont{M.}~\bibnamefont{Troyer}}, \bibnamefont{and}
  \bibinfo{author}{\bibfnamefont{A.~J.} \bibnamefont{Millis}},
  \bibinfo{journal}{Phys. Rev. Lett.} \textbf{\bibinfo{volume}{97}},
  \bibinfo{eid}{076405} (\bibinfo{year}{2006}).

\bibitem[{\citenamefont{Schir\'{o} and Fabrizio}(2009)}]{Keldysh_short}
\bibinfo{author}{\bibfnamefont{M.}~\bibnamefont{Schir\'{o}}} \bibnamefont{and}
  \bibinfo{author}{\bibfnamefont{M.}~\bibnamefont{Fabrizio}},
  \bibinfo{journal}{Phys. Rev. B} \textbf{\bibinfo{volume}{79}},
  \bibinfo{eid}{153302} (\bibinfo{year}{2009}).

\bibitem[{\citenamefont{M\"{u}hlbacher and Rabani}(2008)}]{Rabani}
\bibinfo{author}{\bibfnamefont{L.}~\bibnamefont{M\"{u}hlbacher}}
  \bibnamefont{and} \bibinfo{author}{\bibfnamefont{E.}~\bibnamefont{Rabani}},
  \bibinfo{journal}{Phys. Rev. Lett.} \textbf{\bibinfo{volume}{100}},
  \bibinfo{eid}{176403} (\bibinfo{year}{2008}).

\bibitem[{\citenamefont{Werner et~al.}(2009)\citenamefont{Werner, Oka, and
  Millis}}]{Werner_Keldysh_09}
\bibinfo{author}{\bibfnamefont{P.}~\bibnamefont{Werner}},
  \bibinfo{author}{\bibfnamefont{T.}~\bibnamefont{Oka}}, \bibnamefont{and}
  \bibinfo{author}{\bibfnamefont{A.~J.} \bibnamefont{Millis}},
  \bibinfo{journal}{Phys. Rev. B} \textbf{\bibinfo{volume}{79}},
  \bibinfo{eid}{035320} (\bibinfo{year}{2009}).

\bibitem[{\citenamefont{Kadanoff and Baym}(1962)}]{Kadanoff_Baym}
\bibinfo{author}{\bibfnamefont{L.~P.} \bibnamefont{Kadanoff}} \bibnamefont{and}
  \bibinfo{author}{\bibfnamefont{G.}~\bibnamefont{Baym}},
  \emph{\bibinfo{title}{Quantum Statistical Mechanics}}
  (\bibinfo{publisher}{Benjamin, New York}, \bibinfo{year}{1962}).

\bibitem[{\citenamefont{Rammer}(2007)}]{Rammer_book_noneq}
\bibinfo{author}{\bibfnamefont{J.}~\bibnamefont{Rammer}},
  \emph{\bibinfo{title}{Quantum Field Theory of Nonequilibrium States}}
  (\bibinfo{publisher}{Cambridge University Press}, \bibinfo{year}{2007}).

\bibitem[{\citenamefont{Wagner}(1991)}]{Wagner}
\bibinfo{author}{\bibfnamefont{M.}~\bibnamefont{Wagner}},
  \bibinfo{journal}{Phys. Rev. B} \textbf{\bibinfo{volume}{44}},
  \bibinfo{eid}{6104} (\bibinfo{year}{1991}).

\bibitem[{\citenamefont{Keiter and Morandi}(1984)}]{Keiter_84}
\bibinfo{author}{\bibfnamefont{K.}~\bibnamefont{Keiter}} \bibnamefont{and}
  \bibinfo{author}{\bibfnamefont{G.}~\bibnamefont{Morandi}},
  \bibinfo{journal}{Phys. Rep} \textbf{\bibinfo{volume}{109}},
  \bibinfo{eid}{227} (\bibinfo{year}{1984}).

\bibitem[{\citenamefont{Werner and Millis}(2006)}]{Werner_ctqmc_matrix}
\bibinfo{author}{\bibfnamefont{P.}~\bibnamefont{Werner}} \bibnamefont{and}
  \bibinfo{author}{\bibfnamefont{A.~J.} \bibnamefont{Millis}},
  \bibinfo{journal}{Phys. Rev. B} \textbf{\bibinfo{volume}{74}},
  \bibinfo{eid}{155107} (\bibinfo{year}{2006}).

\bibitem[{\citenamefont{Haule}(2007)}]{Haule_ctqmc}
\bibinfo{author}{\bibfnamefont{K.}~\bibnamefont{Haule}},
  \bibinfo{journal}{Phys. Rev. B} \textbf{\bibinfo{volume}{75}},
  \bibinfo{eid}{155113} (\bibinfo{year}{2007}).

\bibitem[{\citenamefont{Kamenev}(2009)}]{Kamenev}
\bibinfo{author}{\bibfnamefont{A.}~\bibnamefont{Kamenev}},
  \bibinfo{journal}{Advances in Physics} \textbf{\bibinfo{volume}{58}},
  \bibinfo{eid}{197} (\bibinfo{year}{2009}).

\bibitem[{\citenamefont{Prokof'ev and
  Svistunov}(1998)}]{Prokofev_Svistunov_diagMC_98}
\bibinfo{author}{\bibfnamefont{N.~V.} \bibnamefont{Prokof'ev}}
  \bibnamefont{and} \bibinfo{author}{\bibfnamefont{B.~V.}
  \bibnamefont{Svistunov}}, \bibinfo{journal}{Phys. Rev. Lett.}
  \textbf{\bibinfo{volume}{81}}, \bibinfo{pages}{2514} (\bibinfo{year}{1998}).

\bibitem[{\citenamefont{Abrikosov et~al.}(1963)\citenamefont{Abrikosov,
  Dzjaloshinsky, and Gorkov}}]{ADG}
\bibinfo{author}{\bibfnamefont{A.}~\bibnamefont{Abrikosov}},
  \bibinfo{author}{\bibfnamefont{I.}~\bibnamefont{Dzjaloshinsky}},
  \bibnamefont{and} \bibinfo{author}{\bibfnamefont{L.}~\bibnamefont{Gorkov}},
  \emph{\bibinfo{title}{Methods of Quantum Field Theory in Statistical
  Mechanics}} (\bibinfo{publisher}{Dover}, \bibinfo{year}{1963}).

\bibitem[{\citenamefont{Egger et~al.}(2000)\citenamefont{Egger, M\"uhlbacher,
  and Mak}}]{Egger_Mulbacher}
\bibinfo{author}{\bibfnamefont{R.}~\bibnamefont{Egger}},
  \bibinfo{author}{\bibfnamefont{L.}~\bibnamefont{M\"uhlbacher}},
  \bibnamefont{and} \bibinfo{author}{\bibfnamefont{C.~H.} \bibnamefont{Mak}},
  \bibinfo{journal}{Phys. Rev. E} \textbf{\bibinfo{volume}{61}}
  (\bibinfo{year}{2000}).

\bibitem[{\citenamefont{Metropolis et~al.}(1953)\citenamefont{Metropolis,
  Rosenbluth, Rosenbluth, Teller, and Teller}}]{Metropolis}
\bibinfo{author}{\bibfnamefont{N.}~\bibnamefont{Metropolis}},
  \bibinfo{author}{\bibfnamefont{A.~W.} \bibnamefont{Rosenbluth}},
  \bibinfo{author}{\bibfnamefont{M.~N.} \bibnamefont{Rosenbluth}},
  \bibinfo{author}{\bibfnamefont{A.~H.} \bibnamefont{Teller}},
  \bibnamefont{and} \bibinfo{author}{\bibfnamefont{E.}~\bibnamefont{Teller}},
  \bibinfo{journal}{J. Chem. Phys} \textbf{\bibinfo{volume}{21}},
  \bibinfo{eid}{1087} (\bibinfo{year}{1953}).

\bibitem[{\citenamefont{Poteryaev et~al.}(2008)\citenamefont{Poteryaev,
  Ferrero, Georges, and Parcollet}}]{Michel_ctqmc}
\bibinfo{author}{\bibfnamefont{A.~I.} \bibnamefont{Poteryaev}},
  \bibinfo{author}{\bibfnamefont{M.}~\bibnamefont{Ferrero}},
  \bibinfo{author}{\bibfnamefont{A.}~\bibnamefont{Georges}}, \bibnamefont{and}
  \bibinfo{author}{\bibfnamefont{O.}~\bibnamefont{Parcollet}},
  \bibinfo{journal}{Phys. Rev. B} \textbf{\bibinfo{volume}{78}}
  (\bibinfo{year}{2008}).

\bibitem[{\citenamefont{Nozi\'eres and De~Domincis}(1969)}]{XrayEdge}
\bibinfo{author}{\bibfnamefont{P.}~\bibnamefont{Nozi\'eres}} \bibnamefont{and}
  \bibinfo{author}{\bibfnamefont{C.~T.} \bibnamefont{De~Domincis}},
  \bibinfo{journal}{Phys. Rev.} \textbf{\bibinfo{volume}{178}},
  \bibinfo{pages}{1097} (\bibinfo{year}{1969}).

\bibitem[{\citenamefont{Yuval and Anderson}(1970)}]{AY}
\bibinfo{author}{\bibfnamefont{G.}~\bibnamefont{Yuval}} \bibnamefont{and}
  \bibinfo{author}{\bibfnamefont{P.~W.} \bibnamefont{Anderson}},
  \bibinfo{journal}{Phys. Rev. B} \textbf{\bibinfo{volume}{1}},
  \bibinfo{pages}{1522} (\bibinfo{year}{1970}).

\bibitem[{\citenamefont{Nordlander et~al.}(1999)\citenamefont{Nordlander,
  Pustilnik, Meir, Wingreen, and Langreth}}]{Nordlander_99}
\bibinfo{author}{\bibfnamefont{P.}~\bibnamefont{Nordlander}},
  \bibinfo{author}{\bibfnamefont{M.}~\bibnamefont{Pustilnik}},
  \bibinfo{author}{\bibfnamefont{Y.}~\bibnamefont{Meir}},
  \bibinfo{author}{\bibfnamefont{N.~S.} \bibnamefont{Wingreen}},
  \bibnamefont{and} \bibinfo{author}{\bibfnamefont{D.~C.}
  \bibnamefont{Langreth}}, \bibinfo{journal}{Phys. Rev. Lett.}
  \textbf{\bibinfo{volume}{83}}, \bibinfo{pages}{808} (\bibinfo{year}{1999}).

\bibitem[{\citenamefont{Lobaskin and Kehrein}(2005)}]{Kehrein_Lobaskin_05}
\bibinfo{author}{\bibfnamefont{D.}~\bibnamefont{Lobaskin}} \bibnamefont{and}
  \bibinfo{author}{\bibfnamefont{S.}~\bibnamefont{Kehrein}},
  \bibinfo{journal}{Phys. Rev. B} \textbf{\bibinfo{volume}{71}},
  \bibinfo{pages}{193303} (\bibinfo{year}{2005}).

\bibitem[{\citenamefont{Goldhaber-Gordon
  et~al.}(1998)\citenamefont{Goldhaber-Gordon, G\"ores, Kastner, Shtrikman,
  Mahalu, and Meirav}}]{GoldhaberGordon_prl98}
\bibinfo{author}{\bibfnamefont{D.}~\bibnamefont{Goldhaber-Gordon}},
  \bibinfo{author}{\bibfnamefont{J.}~\bibnamefont{G\"ores}},
  \bibinfo{author}{\bibfnamefont{M.~A.} \bibnamefont{Kastner}},
  \bibinfo{author}{\bibfnamefont{H.}~\bibnamefont{Shtrikman}},
  \bibinfo{author}{\bibfnamefont{D.}~\bibnamefont{Mahalu}}, \bibnamefont{and}
  \bibinfo{author}{\bibfnamefont{U.}~\bibnamefont{Meirav}},
  \bibinfo{journal}{Phys. Rev. Lett.} \textbf{\bibinfo{volume}{81}},
  \bibinfo{pages}{5225} (\bibinfo{year}{1998}).

\bibitem[{\citenamefont{Doyon and Andrei}(2006)}]{Andrei_Doyon_PRB}
\bibinfo{author}{\bibfnamefont{B.}~\bibnamefont{Doyon}} \bibnamefont{and}
  \bibinfo{author}{\bibfnamefont{N.}~\bibnamefont{Andrei}},
  \bibinfo{journal}{Phys. Rev. B} \textbf{\bibinfo{volume}{73}},
  \bibinfo{eid}{245326} (\bibinfo{year}{2006}).

\bibitem[{\citenamefont{Türeci et~al.}(2009)\citenamefont{Türeci, Hanl,
  Claassen, Weichselbaum, Hecht, Braunecker, Govorov, Glazman, von Delft, and
  Imamoglu}}]{Glazman_quench}
\bibinfo{author}{\bibfnamefont{H.~E.} \bibnamefont{Türeci}},
  \bibinfo{author}{\bibfnamefont{M.}~\bibnamefont{Hanl}},
  \bibinfo{author}{\bibfnamefont{M.}~\bibnamefont{Claassen}},
  \bibinfo{author}{\bibfnamefont{A.}~\bibnamefont{Weichselbaum}},
  \bibinfo{author}{\bibfnamefont{T.}~\bibnamefont{Hecht}},
  \bibinfo{author}{\bibfnamefont{B.}~\bibnamefont{Braunecker}},
  \bibinfo{author}{\bibfnamefont{A.}~\bibnamefont{Govorov}},
  \bibinfo{author}{\bibfnamefont{L.}~\bibnamefont{Glazman}},
  \bibinfo{author}{\bibfnamefont{J.}~\bibnamefont{von Delft}},
  \bibnamefont{and} \bibinfo{author}{\bibfnamefont{A.}~\bibnamefont{Imamoglu}},
  \bibinfo{journal}{arXiv:0907.3854}  (\bibinfo{year}{2009}).

\bibitem[{\citenamefont{Silva}(2008)}]{Silva_work_statistics}
\bibinfo{author}{\bibfnamefont{A.}~\bibnamefont{Silva}},
  \bibinfo{journal}{Phys. Rev. Lett.} \textbf{\bibinfo{volume}{101}},
  \bibinfo{eid}{120603} (\bibinfo{year}{2008}).

\bibitem[{\citenamefont{Anderson}(1961)}]{AIM_meanfield}
\bibinfo{author}{\bibfnamefont{P.~W.} \bibnamefont{Anderson}},
  \bibinfo{journal}{Phys. Rev.} \textbf{\bibinfo{volume}{124}},
  \bibinfo{pages}{41} (\bibinfo{year}{1961}).

\bibitem[{\citenamefont{Mahan}(1990)}]{Mahan_book}
\bibinfo{author}{\bibfnamefont{G.}~\bibnamefont{Mahan}},
  \emph{\bibinfo{title}{Many-Particle Physics}} (\bibinfo{publisher}{Plenum
  Press}, \bibinfo{year}{1990}).

\bibitem[{\citenamefont{Caldeira and Leggett}(1981)}]{Caldeira_Leggett_PRL81}
\bibinfo{author}{\bibfnamefont{A.~O.} \bibnamefont{Caldeira}} \bibnamefont{and}
  \bibinfo{author}{\bibfnamefont{A.~J.} \bibnamefont{Leggett}},
  \bibinfo{journal}{Phys. Rev. Lett.} \textbf{\bibinfo{volume}{46}},
  \bibinfo{pages}{211} (\bibinfo{year}{1981}).

\bibitem[{\citenamefont{Withoff and Fradkin}(1990)}]{Fradkin_PRL}
\bibinfo{author}{\bibfnamefont{D.}~\bibnamefont{Withoff}} \bibnamefont{and}
  \bibinfo{author}{\bibfnamefont{E.}~\bibnamefont{Fradkin}},
  \bibinfo{journal}{Phys. Rev. Lett.} \textbf{\bibinfo{volume}{64}},
  \bibinfo{pages}{1835} (\bibinfo{year}{1990}).

\bibitem[{\citenamefont{Gonzalez-Buxton and Ingersent}(1998)}]{Ingersent_PRB}
\bibinfo{author}{\bibfnamefont{C.}~\bibnamefont{Gonzalez-Buxton}}
  \bibnamefont{and}
  \bibinfo{author}{\bibfnamefont{K.}~\bibnamefont{Ingersent}},
  \bibinfo{journal}{Phys. Rev. B} \textbf{\bibinfo{volume}{57}},
  \bibinfo{pages}{14254} (\bibinfo{year}{1998}).

\bibitem[{\citenamefont{Cornaglia et~al.}(2009)\citenamefont{Cornaglia, Usaj,
  and Balseiro}}]{Cornaglia_PRL}
\bibinfo{author}{\bibfnamefont{P.~S.} \bibnamefont{Cornaglia}},
  \bibinfo{author}{\bibfnamefont{G.}~\bibnamefont{Usaj}}, \bibnamefont{and}
  \bibinfo{author}{\bibfnamefont{C.~A.} \bibnamefont{Balseiro}},
  \bibinfo{journal}{Phys. Rev. Lett.} \textbf{\bibinfo{volume}{102}}
  (\bibinfo{year}{2009}).

\bibitem[{\citenamefont{Neto et~al.}(2009)\citenamefont{Neto, Kotov, Nilsson,
  Pereira, Peres, and Uchoa}}]{CastroNeto}
\bibinfo{author}{\bibfnamefont{A.~C.} \bibnamefont{Neto}},
  \bibinfo{author}{\bibfnamefont{V.}~\bibnamefont{Kotov}},
  \bibinfo{author}{\bibfnamefont{J.}~\bibnamefont{Nilsson}},
  \bibinfo{author}{\bibfnamefont{V.}~\bibnamefont{Pereira}},
  \bibinfo{author}{\bibfnamefont{N.}~\bibnamefont{Peres}}, \bibnamefont{and}
  \bibinfo{author}{\bibfnamefont{B.}~\bibnamefont{Uchoa}},
  \bibinfo{journal}{Solid State Comm.} \textbf{\bibinfo{volume}{149}},
  \bibinfo{pages}{1094 } (\bibinfo{year}{2009}).

\bibitem[{\citenamefont{Chen and Jayaprakash}(1998)}]{Chen_PRB}
\bibinfo{author}{\bibfnamefont{K.}~\bibnamefont{Chen}} \bibnamefont{and}
  \bibinfo{author}{\bibfnamefont{C.}~\bibnamefont{Jayaprakash}},
  \bibinfo{journal}{Phys. Rev. B} \textbf{\bibinfo{volume}{57}},
  \bibinfo{pages}{5225} (\bibinfo{year}{1998}).

\bibitem[{\citenamefont{Galpin and Logan}(2008)}]{Logan_PRB}
\bibinfo{author}{\bibfnamefont{M.~R.} \bibnamefont{Galpin}} \bibnamefont{and}
  \bibinfo{author}{\bibfnamefont{D.~E.} \bibnamefont{Logan}},
  \bibinfo{journal}{Phys. Rev. B} \textbf{\bibinfo{volume}{77}},
  \bibinfo{pages}{195108} (\bibinfo{year}{2008}).

\bibitem[{\citenamefont{Anders et~al.}(2007)\citenamefont{Anders, Bulla, and
  Vojta}}]{Anders_Vojta_spinboson}
\bibinfo{author}{\bibfnamefont{F.~B.} \bibnamefont{Anders}},
  \bibinfo{author}{\bibfnamefont{R.}~\bibnamefont{Bulla}}, \bibnamefont{and}
  \bibinfo{author}{\bibfnamefont{M.}~\bibnamefont{Vojta}},
  \bibinfo{journal}{Phys. Rev. Lett.} \textbf{\bibinfo{volume}{98}},
  \bibinfo{pages}{210402} (\bibinfo{year}{2007}).

\bibitem[{\citenamefont{Gull}(2008)}]{Gull_PhD}
\bibinfo{author}{\bibfnamefont{E.}~\bibnamefont{Gull}},
  \emph{\bibinfo{title}{Continuous-Time Quantum Monte Carlo Algorithms for
  Fermions}} (\bibinfo{publisher}{PhD Thesis, ETH Zurich},
  \bibinfo{year}{2008}).

\bibitem[{\citenamefont{Prokof'ev and Svistunov}(2007)}]{Prokofev_sign}
\bibinfo{author}{\bibfnamefont{N.}~\bibnamefont{Prokof'ev}} \bibnamefont{and}
  \bibinfo{author}{\bibfnamefont{B.}~\bibnamefont{Svistunov}},
  \bibinfo{journal}{Phys. Rev. Lett.} \textbf{\bibinfo{volume}{99}},
  \bibinfo{eid}{250201} (\bibinfo{year}{2007}).

\end{thebibliography}

\end{document}